%% file: main.tex
\documentclass[conference]{IEEETran}
\usepackage{booktabs} 
\usepackage{placeins}
\usepackage{epsfig,endnotes}
\usepackage[tight,footnotesize]{subfigure}
\usepackage{mathtools}
\usepackage{graphicx,xcolor}
\usepackage{epstopdf}
\usepackage{array}
\usepackage{color}
\usepackage{algpseudocode}
\usepackage[english]{babel}
\usepackage[utf8]{inputenc}
\usepackage[colorinlistoftodos]{todonotes}
\usepackage{algpseudocode}
\usepackage{multirow}
\usepackage{tabularx}
\usepackage{array}
\usepackage{setspace}
\usepackage{textcomp}
\usepackage{verbatim}
\usepackage{xspace}
\usepackage{threeparttable}
\usepackage[ruled, linesnumbered, vlined]{algorithm2e}
\usepackage[font={bf},textfont=bf]{caption}
\usepackage{hyperref}  
\usepackage{todonotes}
\usepackage{ulem}  
\usepackage{bbding}
\usepackage{fancyhdr}
\usepackage{diagbox} 
\usepackage{float}
\usepackage{pifont} 
\usepackage{colortbl}
\usepackage{threeparttable} 
\usepackage[citestyle=numeric-comp, backend=bibtex, sorting=none]{biblatex}
\renewbibmacro{in:}{} 
\usepackage{titlesec}
\usepackage{enumitem}

\newcommand{\blue}[1]{\textcolor{black}{#1}} 

\newcommand{\alias}{\texttt{PowerRadio}\xspace} 
\newcommand{\fig}[1]{Fig.~\ref{#1}} 
\newcommand{\tab}[1]{Tab.~\ref{#1}}
\newcommand{\eq}[1]{Eq.~\eqref{#1}}
\newcommand{\mysec}[1]{Sec.~\ref{#1}}

\bibliography{mybib} 

\ifCLASSINFOpdf
\else
\fi
\begin{document}
%
\title{\alias: Manipulate Sensor Measurement\\ via Power GND Radiation}

	


\author{\IEEEauthorblockN{Yan Jiang\textsuperscript{1},
Xiaoyu Ji\textsuperscript{1\ding{41}}\thanks{\textsuperscript{\ding{41}} Xiaoyu Ji is the corresponding author},
Yancheng Jiang\textsuperscript{1}, 
Kai Wang\textsuperscript{1},
Chenren Xu\textsuperscript{2} and
Wenyuan Xu\textsuperscript{1}}
\IEEEauthorblockA{\textsuperscript{1}Zhejiang University, China, \{yj98, xji, jiangyancheng, eekaiwang, wyxu\}@zju.edu.cn}
\IEEEauthorblockA{\textsuperscript{2}Peking University, China, chenren@pku.edu,cn}}



\graphicspath{{figures/}}

\IEEEoverridecommandlockouts
\makeatletter\def\@IEEEpubidpullup{6.5\baselineskip}\makeatother
\IEEEpubid{\parbox{\columnwidth}{
		Network and Distributed System Security (NDSS) Symposium 2025\\
		23–28 February 2025, San Diego, CA, USA\\
		ISBN 979-8-9894372-8-3\\
		https://dx.doi.org/10.14722/ndss.2025.230295\\
		www.ndss-symposium.org
}
\hspace{\columnsep}\makebox[\columnwidth]{}}

\maketitle



%
\IEEEpeerreviewmaketitle
\input{sections/Abstract}

\input{sections/Introduction}

\input{sections/Background}
\input{sections/ThreatModel}

\input{sections/Feasibility}

\input{sections/Design}

\input{sections/Evaluation}

\input{sections/RelatedWork}

\input{sections/Discussion}
\input{sections/Conclusion}

\input{sections/Acknowlegment}
    \printbibliography 
\input{sections/Appendix}
\end{document}

%% file: sections/Abstract.tex

\begin{abstract}\label{sec: Abstract}Sensors are key components enabling various applications, e.g., home intrusion detection and environmental monitoring. While various software defenses and physical protections are used to prevent sensor manipulation, this paper introduces a new threat vector,~\alias, that bypasses existing protections and changes sensor readings from a distance.~\alias leverages interconnected ground (GND) wires, a standard practice for electrical safety at home, to inject malicious signals. The injected signal is coupled by the sensor's analog measurement wire and eventually survives the noise filters, inducing incorrect measurement. We present three methods to manipulate sensors by inducing static bias, periodical signals, or pulses. For instance, we show adding stripes into the captured images of a surveillance camera or injecting inaudible voice commands into conference microphones. We study the underlying principles of~\alias and identify its root causes: (1) the lack of shielding between ground and data signal wires and (2) the asymmetry of circuit impedance that enables interference to bypass filtering.
We validate~\alias against a surveillance system, broadcast systems, and various sensors.
We believe that~\alias represents an emerging threat, exhibiting the advantages of both radiated and conducted EMI, e.g., expanding the effective attack distance of radiated EMI yet eliminating the requirement of line-of-sight or approaching physically. Our insights shall provide guidance for enhancing the sensors' security and power wiring during the design phases.
\end{abstract}

%% file: sections/Introduction.tex
\section{Introduction}
\label{sec: introduction}


Sensors are key components widely used in various application scenarios. For example, a surveillance camera in a smart home helps homeowners identify intruders or monitor suspicious activities. Securing sensors is therefore for protecting user safety and privacy. Various studies have been conducted to prevent sensor manipulation, including software defenses (e.g., false data detection) and physical protections (e.g., voltage regulators, and EM shielding). However, we identify a new attack vector against sensors and design~\alias, which bypasses those protection methods and changes sensor readings from a distance.

\alias leverages the interconnected ground (GND) wires, a standard practice for electrical safety at home~\cite{elie2024houseearth}, to inject malicious signals. Specifically, the injected signal couples with the sensor's analog measurement wire, survives the noise filters, and induces common mode (CM) current, which is converted into a differential mode (DM) signal that ultimately affects the sensor reading. We studied the underlying principles of~\alias and identified its root causes: the lack of shielding between ground and data signal wires and the asymmetry of circuit impedance, which enables interference to bypass filtering and induce false sensor measurements.
\begin{figure}[!t]  
    \centering
    \includegraphics[width=1\linewidth]{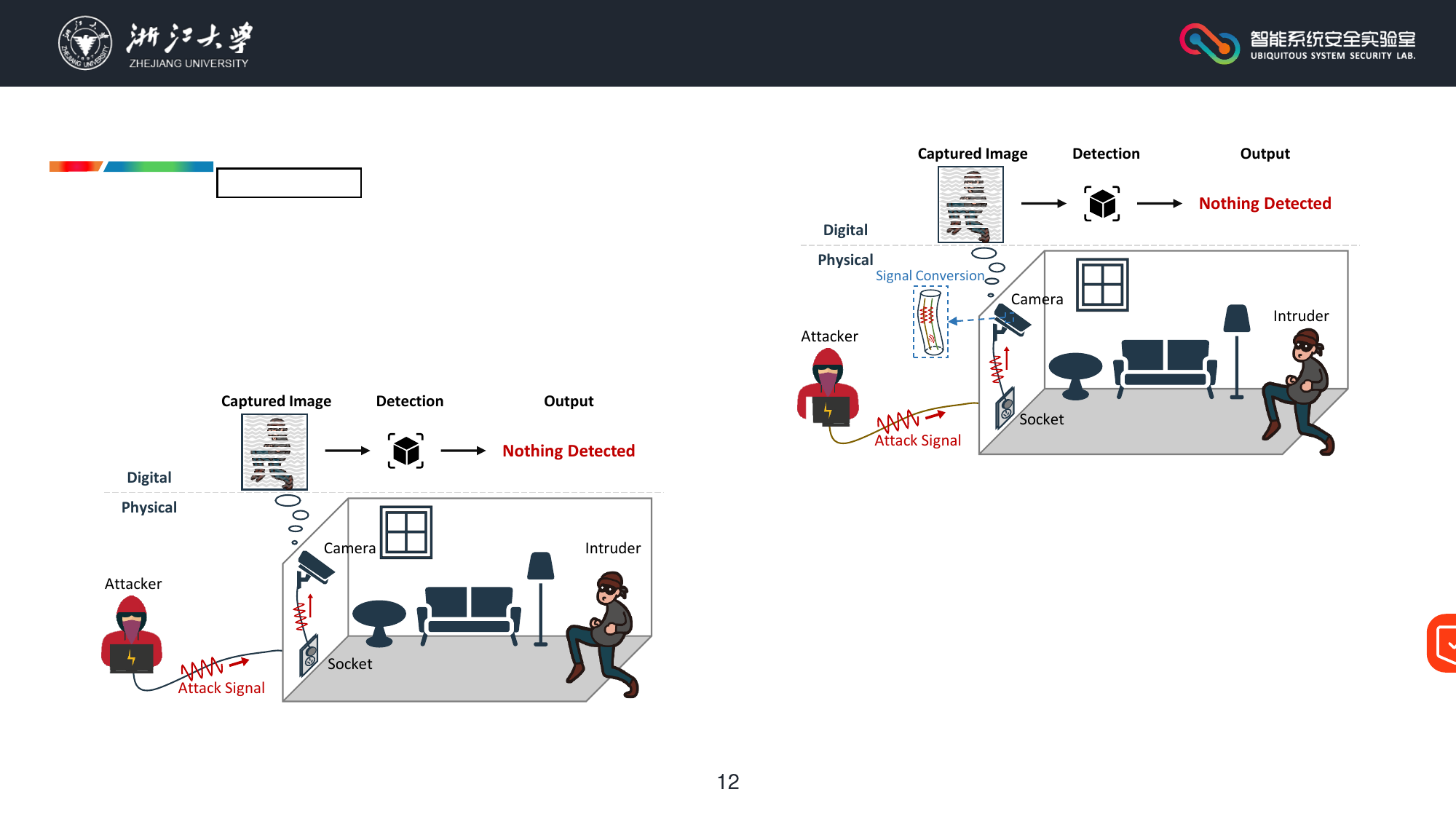}
    \caption{Illustration of~\alias in an intrusion detection scenario. An attacker injects a crafted signal into the GND port of a public charging socket. The signal travels through the GND wiring to the surveillance camera, inducing stripes on the captured image that bypass the detection algorithm. As a result, the intrusion detection system fails to detect the burglar and trigger an alert, compromising the homeowner's safety.}
    \label{fig: overview}
\end{figure}
Moreover, we uncover three key factors that influence the asymmetry of circuit impedance and validate our analysis through simulations and experiments. Additionally, we explore vulnerabilities in signal processing modules, such as the nonlinearity of an amplifier, and present three signal design methods to manipulate sensors by inducing static bias, periodical signals, and pulses. For instance, we demonstrate injecting inaudible voice commands into conference microphones or false pulses into a speed sensor. 

While an attacker generally cannot physically approach or modify the firmware of the target sensor, they can use~\alias to achieve a cross-socket or even cross-room manipulation attack on the indoor sensor. For instance,~\fig{fig: overview} illustrates a home intrusion detection scenario where a surveillance camera captures the home scene, identifies intruders, and triggers an alert to the homeowner. However, we demonstrate that an attacker can evade the detection algorithm by injecting a crafted signal into the GND wire of the outdoor wall socket, introducing stripes into the captured images that prevent the camera from detecting intruder.

Unlike existing physical signal-based injection attacks, which manipulate sensors using either wireless signals such as EMI~\cite{wang2022ghosttouch, tu2019trick}, laser~\cite{sugawara2020light, jin2023pla}, sound~\cite{ji2021poltergeist, bolton2018bluenote}, ultrasound~\cite{zhang2017dolphinattack}, or magnetic fields~\cite{barua2020hall}, or wired signals like power supply voltages~\cite{wang2023volttack}, we introduce a new attack vector,~\alias, that uses GND wires to transmit and radiate the attack signal. 
We believe~\alias represents an emerging threat that combines the advantages of both radiated and conducted EMI. For instance, the prior study~\cite{wang2023volttack} requires a more powerful attacker capable of compromising the power source, and its injected differential mode voltages are easily mitigated by noise filters and voltage stabilizers. Instead,~\alias requires only access the GND wire and focuses on transmitting common mode voltage with an almost infinite load impedance, enabling it to bypass existing defense methods. In addition,~\alias extends the effective attack distance of radiated EMI yet eliminates the requirement for line-of-sight or physical proximity. For instance,~\alias can interfere with image and microphone sensors up to 15\,m, while previous wireless EMI works~\cite{kohler2022signal, esteves2018remote} targeted image sensors at 50\,cm and microphone sensors at 3\,m. 
Furthermore, our insight into~\alias shall provide guidance for enhancing sensor security and power wiring design, such as integrating an attack detection module into the power wiring or optimizing circuit asymmetry to improve protection.



However, to achieve~\alias, we have to overcome the following two critical challenges:

(1) \textit{How to effectively inject attack signals?} 
Sensors are known to be affected by the power supply, therefore, various prevention methods, such as voltage regulators and noise filters, have been employed. 
As a result, attack signals injected into the power cable are treated as noises and are eliminated. 
To avoid being filtered and prevent damage to the hardware of other devices connected to the same power network, 
we present a new attack method that bypasses voltage regulators and effectively injects attack signals into the target sensor from a distance. Specifically, we inject attack signals into the GND wire, allowing the transmission of the signal without being filtered, since there are no suspicious voltage fluctuations between the live wire and the GND wire.
Furthermore, the GND wire can function as an antenna, radiating the attack signal to the sensor's analog signal, thus altering the sensor readings. Thus, we term this attack~\alias. Our investigation shows that the fundamental concept underpinning this attack is signal coupling and conversion. 
Hence, for an attacker to effectively inject attack signals into the target sensor, optimizing signal frequencies and amplitudes is crucial for enhancing coupling and conversion efficiency.


(2) \textit{How to create a given output?} 
Given the limited control an attacker has over shaping the injected signal within the sensor, achieving a precise output for the target sensor poses a significant challenging. To address this challenge, we investigate vulnerabilities in critical signal processing circuits of sensors, inducing the nonlinearity of amplifiers (AMPs), the over-sensitivity of comparators (CMPs), and the sampling distortion of analog-to-digital converters (ADCs). By leveraging these vulnerabilities, we propose three signal-crafting techniques: (1) modulation-based alternating current (AC) injection, (2) jitter-based pulse injection, (3) and biasing-based direct current (DC) injection. These techniques enable the induction of malicious AC, pulse, and DC in sensors, respectively. Essentially, an attacker initially analyzes vulnerable signal processing circuits and the legitimate outputs of target sensors. Subsequently, the attacker uses a specific output as the baseband signal and selects an appropriate signal design technique. This approach allows the attacker to generate a finely adjustable output that mimics the legitimate output, thereby reducing the risk of detection. 



In summary, our contributions are summarized as follows:
\begin{itemize}[leftmargin=6mm] 
    \item We present \alias, a new physical attack method for sensor manipulation. This attack enables an adversary to bypass software and physical defenses, altering sensor readings from a distance, even across rooms or through wall sockets.
    \item  We identify an emerging threat, using the interconnected GND wire to transmit inference. This approach combines the advantages of both radiated and conducted EMI, such as extending the effective attack distance of radiated EMI while eliminating the need for line-of-sight or physical proximity. 
    \item We systematically analyze the underlying principles of energy conversion and validate our findings through modeling, simulations, and experiments. Furthermore, our insights shall guide for enhancing the sensor security and power wiring design during the development phase.
    \item We evaluate~\alias on 17 sensors, 2 commercial systems, validate~\alias's effectiveness in a home wiring scenario, and propose potential countermeasures to mitigate this threat.
\end{itemize}

%% file: sections/Background.tex
\section{Background}
\label{sec: background}

\subsection{Sensors and electronic modules}\label{sec: back_iot_sensor_module}


A sensor is used to detect physical stimuli, such as light or motion, in the real world and convert them into electrical signals~\cite{fraden1994handbook}. 
A sensor typically consists of two parts: (a) \textit{Transducer:} which measures a physical stimulus and produces an analog electrical representation; (b) \textit{Signal processing circuit:} which reduces noise and amplifies useful information, incorporating components such as amplifiers and comparators. 
Sensors are widely used in applications such as smart homes and industrial systems. Therefore, securing them is crucial for protecting user's safety and privacy. 

Electronic modules refer to discrete components or integrated circuits that perform specific functions~\cite{laughton2013electrical,tietze2015electronic}, \blue{such as amplifiers (AMPs) and analog-to-digital converters (ADCs).}
For instance, an ADC converts analog signals into digital signals, while a digital-to-analog converter (DAC) performs the reverse function.

\subsection{CM and DM signals}

Electromagnetic interference (EMI) refers to unwanted electromagnetic disturbances or noise signals that affect electronic systems~\cite{redoute2009emc}. 
Based on the signal type, EMI can be classified into two categories: Differential Mode (DM) signals and Common Mode (CM) signals. DM signals travel in opposite directions within a pair of transmission lines, and the DM voltage is defined as the voltage difference between the two conductors. Conversely, CM signals of equal magnitude flow in the same direction along the transmission line and are intended to traverse the referenced GND due to parasitic capacitors~\cite{wang2010investigation}. In an ideal scenario, pure CM signals, upon reaching termination, result in zero DM voltage and current. Thus, CM signals theoretically do not impact the operation of system elements. 
However, in practice, electrical systems with imperfectly symmetric structures induce energy conversion between DM and CM signals~\cite{wang2010investigation, davide2017modeling, jaze2013differential}, which interferes with voltage measurements. 

\subsection{Grounding}\label{sec: grounding}
Grounding plays a critical role in maintaining signal integrity and mitigating interference in electronic components. In this work, we leverage interconnected GND wires to transmit attack signals and bypass filtering emchanisms. GND is typically categorized into the following types: \textit{(1) Signal GND:} is defined as a common reference point in a circuit from which voltages are measured and provides a low-impedance path for currents. \textit{(2) Chassis GND:} refers to the connection of components to the metal frame (chassis) of equipment, such as vehicles, ensuring the external surfaces of the device remain at the same electrical potential as the reference GND. 
\textit{(3) Earth GND:} refers to a physical connection to the Earth, serving as a common reference potential for the device and providing an absolute GND reference. Generally, GNDs are interconnected to protect internal circuits from surges and shield against external EMI.

%% file: sections/ThreatModel.tex

\section{\ Threat Model}
\label{sec: threatmodel}
\begin{itemize}[leftmargin=5mm]
    \item \textbf{Attacker's Goal.} The attacker aims to induce false measurements for sensors by injecting a crafted signal into a power cable without accessing the data transmission or directly altering the sensed environment.
    
    \item \textbf{Capability and Knowledge.} The attacker cannot access the sensor except its GND cable. For example, an attacker can gain access to an indoor sensor's GND by leveraging the home's interconnected GND wiring via the outdoor wall socket. This assumption is reasonable, as every dwelling unit is typically required to have at least one outdoor wall socket each at both the front and rear of the house~\cite{nec2015general}, thereby providing attackers with access to the GND. Furthermore, we assume that the attacker knows the target sensor's model through social engineering methods, such as shoulder surfing, referencing online device datasheets (official or unofficial), or examining device teardown videos or reports. The attacker can also acquire a similar sensor for assessment beforehand. 

    
    \item \textbf{Attack Device.} To stealthily implement~\alias (i.e., establish contact with the GND of the target device), we assume that the attacker can install~\alias behind a wall as a power plug, as shown in~\fig{fig: overview}, or disguise the attack device as a power station or charging device, e.g., a desktop computer, which is plugged into the wall socket. 
    Alternatively, it's feasible to downsize the attack device using a development board (e.g., AD2~\cite{AnalogDiscovery2}) and an integrated power amplifier module (e.g., ATA-M230~\cite{ATA-M230}) for future research. This combination can provide a high voltage of 600\,Vpp at 700\,kHz, exceeding the current requirements of 300\,Vpp at 500\,kHz.

\end{itemize}

%% file: sections/Feasibility.tex

\section{\ Principle of Attacks}
\label{sec: feasibility}
\begin{figure*}[!t]  
	\centering
	\includegraphics[width=1\linewidth]{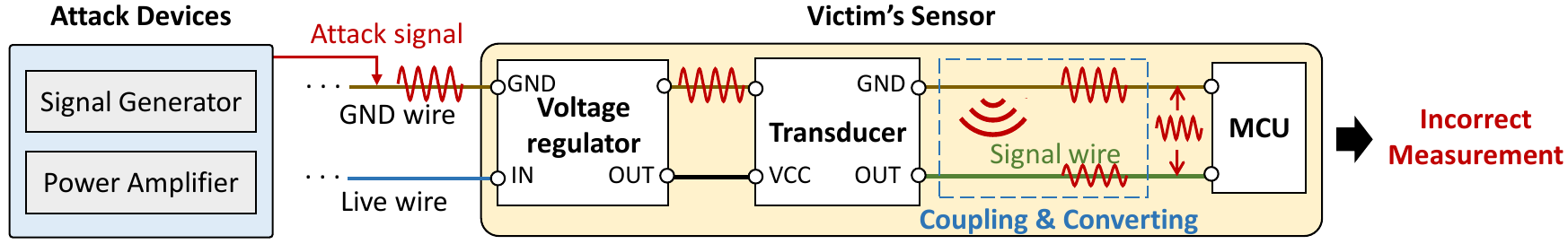}
	\caption{Work flow of~\alias. The attack signal is injected into a GND wire connected to the target sensor's GND. Then, the signal bypasses the voltage regulator because there is no voltage difference between the live and the GND wire. Subsequently, the GND wire acts as an antenna, radiating the attack signal to the analog signal wire through parasitic capacitors and inducing CM current. Finally, the electronic imbalance converts the CM current into a DM voltage at the MCU input, resulting in inaccurate measurements.}   
	\label{fig: principle_main}
\end{figure*}

In this section, we will elaborate on the fundamental principles of~\alias.

\subsection{Energy Conversion Modeling} \label{sec: energy_conversion_model}
\begin{figure} [!t]  
	\centering
        \subfigure[Coupling Model.]{
		\includegraphics[height=0.31\linewidth]{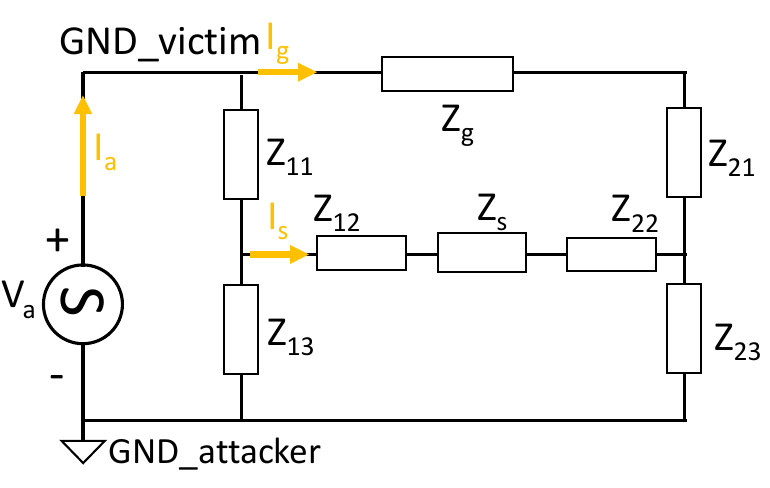}
		\label{fig: principle_coupling2}
	}
        \hspace{-4mm}
	\subfigure[Converting Model.]{
		\includegraphics[height=0.31\linewidth]{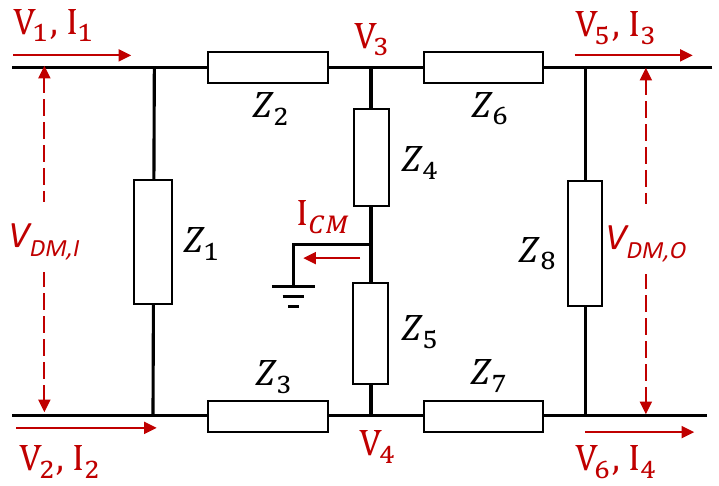}
		\label{fig: principle_conversion2}
	}
    \vspace{-2pt}
	\caption{Illustration of equivalent circuit diagrams for simplified models after delta-Y transformations. (a) Coupling model: The attack signal is coupled from the GND wire to the signal wire, inducing a CM current. (b) Converting model: The CM current is converted into a DM voltage.}\label{fig: circuit}
    \vspace{-4pt}
\end{figure}

Our study identifies two critical parallel energy-convertion stages in~\alias attacks: the coupling stage and the converting stage, as shown in~\fig{fig: principle_main}. To illustrate this, we utilize simple load resistances and dependent sources to establish two energy conversion models, as depicted in~\fig{fig: circuit}. 

\textbf{Coupling Stage.} In the coupling stage, the GND wire acts as a potential antenna, emitting radiated electromagnetic energy to the nearby parallel signal wire through parasitic capacitors~\cite{pcbdesignTI, zumbahlen2011linear}. 

To simplify the circuit, we convert the parasitic capacitance into impedance and perform a delta-y transformation~\cite{laughton2013electrical} as shown in~\fig{fig: principle_coupling2}. Here, $V_{a}$ represents the attack signal, $Z_g$ and $Z_s$ represent the line impedance of the victim device's GND line and the signal line, $Z_{ij}$ refers to equivalent impedance of parasitic capacitors, loads and transmission lines, $GND\_{victim}$ and $GND\_{attacker}$ denote the referenced signal GND wire of the victim device and attack device, respectively. Within~\fig{fig: principle_coupling2}, there are three closed current loops, where $I_a$ denotes the current flowing in the signal line of the attack source, while $I_g$ and \blue{$I_s$} represent the propagating currents along the GND line and the signal line of the victim device, respectively. By applying Kirchhoff's voltage law (KVL) to these three current loops, we obtain: 
\begin{equation}
\begin{aligned}
    Z_{11}(I_a-I_g) + Z_{13}(I_a-I_g-I_s) -V_s &= 0\\[2pt]
    Z_{11}(I_a-I_g) + (Z_{12}+Z{v}+Z_{22})I_s-Z_{21}I_g-Z_gI_g &= 0\\[2pt]
    (Z_{12}+Z{v}+Z_{22})I_s + Z_{23}(I_s+I_g)-Z_{13}(I_a-I_g-I_v) &= 0\label{con: cp}
\end{aligned}
\end{equation}



We define the CM current induced by the attack signal as the average current flowing through both the GND line and the signal line of the victim device, given by:
\begin{gather}
\setlength{\abovedisplayskip}{1pt}
\setlength{\belowdisplayskip}{1pt}
    I_{CM}=\frac{I_g+I_s}{2}=\mu V_s\label{eq: ICM_Vs}
\end{gather}
where the coefficient $\mu$, representing the coupling factor, is determined according to~\eq{con: cp} 
Essentially, a larger $\mu$ corresponds to a greater CM current on the transmission lines. Essentially, a larger $\mu$ corresponds to a greater CM current on the transmission lines.

\textbf{Converting Stage.} 
Another parallel stage is the converting stage, where the CM current transforms into a DM voltage due to the asymmetry of the electric circuits, ultimately inducing false measurements in the sensor. 
Similar to the coupling stage, we use simple impedance and transmission lines to depict the CM-DM conversion model and adopt a group of polynomials to quantize the affecting factor, i.e., the imbalance factor, and perform a delta-y transformation~\cite{laughton2013electrical} to simplify 
The resulting simplified circuit is shown in~\fig{fig: principle_conversion2}, where $I_1$, $I_2$, $V_1$, and $V_2$ represent the input voltages and currents over transmission lines, respectively. $I_3$, $I_4$, $V_5$, and $V_6$ represent the output voltages and currents over transmission lines, respectively. 
It should be noted that the $I_1$ and $I_2$ are not equal to $I_3$ and $I_4$, because the coupling model is considered a closed circuit, whereas the conversion model depicts describes open transmission lines, including input current and output current. We define the input DM voltage $V_{DM, I}$ as the voltage difference between two transmission lines on the input side, i.e., $V_{DM, I} = V_1-V_2$, and the input CM current $I_{CM}$ as the total current that flows on both transmission lines, i.e., $I_{CM} = I_1+I_2-I_3-I_4$. Besides, according to Kirchhoff's laws~\cite{laughton2013electrical}, we have:
\begin{equation}
\begin{aligned}
I_1 - \frac{V_1-V_3}{Z_2} - \frac{V_1-V_2}{Z_1} &= 0 \\[2pt]
I_2 + \frac{V_1-V_2}{Z_1} + \frac{V_4-V_2}{Z_3} &= 0 \\[2pt]
\frac{V_1-V_3}{Z_2} - \frac{V_3-V_5}{Z_6}- \frac{V_3}{Z_4} &= 0\\[2pt]
\frac{V_6-V_4}{Z_7} - \frac{V_4-V_2}{Z_3} - \frac{V_4}{Z_5} &= 0\\[2pt]
\frac{V_3-V_5}{Z_6} - \frac{V_5-V_6}{Z_8} &= I_3\\[2pt]
\frac{V_5-V_6}{Z_8} - \frac{V_6-V_4}{Z_7} &= I_4
\end{aligned}
\end{equation}
Assume the output voltage $ V_{DM, O} = V_5- V_6$ represents the output of the victim's device, i.e., the voltage difference across $Z_8$. Combining the equations above yields the result for $ V_{DM, O}$,
\begin{gather}
\setlength{\abovedisplayskip}{1pt}
\setlength{\belowdisplayskip}{1pt}
    V_{DM,O} = k_1 V_{DM,I}+k_2 I_{CM}+k_3 I_3+k_4 I_4\label{con: define_vdmo}
\end{gather}
where $k_1$, $k_2$, $k_3$ and $k_4$ are constant coefficients determined by the impedance. The equation \eq{con: define_vdmo} demonstrates that the output voltage $V_{DM, O}$ is a combination of four components: the input DM voltage $V_{DM, I}$, the input CM current $I_{CM}$, the output currents $I_3$ and $I_4$. Consequently, the coefficient $k_2$ represents the degree of CM-DM conversion and can be expressed as:
\begin{equation}
    \begin{aligned}
       k_2 &= c_1c_2(h_1+h_2)\\[2pt]
       h_1 &= Z_R(Z_{1O}-Z_{2O})\\[2pt]
       h_2 &= Z_{2O}(Z_L-Z_R)
    \end{aligned}
\end{equation}
where $c_1$ and $c_2$ are constants, and $h_1$ and $h_2$ are composed of parasitic impedance and line impedance of the circuit, respectively. These parameters represent the degree of asymmetry and are thus termed as the 
\textit{asymmetric factor}. Upon examining the expressions of $h_1$ and $h_2$, it becomes evident that CM-DM conversion will not occur only if both the parasitic impedance is perfectly symmetrical (i.e., $Z_{1O}=Z_{2O}$) and the line impedance is perfectly symmetrical (i.e., $Z_R=Z_L$). However, achieving perfect symmetry in practical designs is challenging, making it difficult to entirely eliminate CM-to-DM conversion. 

\begin{figure} [!t]  
	\centering
	\setlength{\abovecaptionskip}{-0.05cm}
	\subfigure[FRC of $I_{CM}$.]{
		\includegraphics[height=0.3\linewidth]{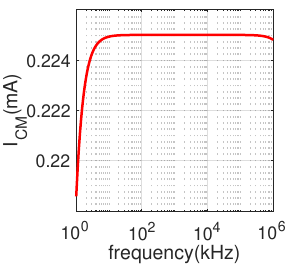}		\label{fig: principle sim ICM}
	}
        \hspace{-5mm} 
	\subfigure[FRC of $k_2$.]{
		\includegraphics[height=0.3\linewidth]{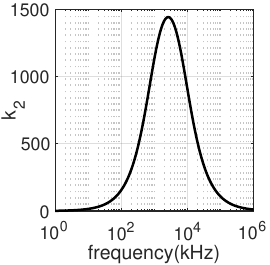}
		\label{fig: principle sim k2}
	}
        \hspace{-5mm}
        \subfigure[FRC of $V_{DM}$.]{
		\includegraphics[height=0.3\linewidth]{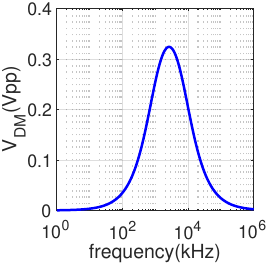}
		\label{fig: principle sim VDM}
	}
	\caption{Simulation results of coupling and conversion models. (a) FRC of CM current $I_{CM}$. (b) FRC of conversion coefficient $k_2$. (c) FRC of DM output voltage $V_{DM}$.}
	\label{fig: CMDMresults}
\end{figure}

\textbf{Simulation Analysis.} To validate our analysis, we used MATLAB~\cite{mathwork2024matlab} for simulation analysis. Our simulations involved two circuits based on~\fig{fig: principle_coupling2} and~\fig{fig: principle_conversion2}, enabling us to quantitatively explore the frequency response curve (FRC) for CM current $I_{CM}$, the CM-DM conversion coefficient $k_2$, and the output DM voltages $V_{DM}$ during an~\alias attack. Based on prior research\cite{TI2012Design, analogXXXXPCBDesign, Pajovic2007Capacitance, wang2004parasitic}, the typical resistance and parasitic capacitance between two transmission lines on PCB boards range from $1\sim100M\Omega$ and $0.1\sim10\mu F$, respectively. Detailed simulation parameters are provided in Appendix~\ref{sec: appendix_modeling_simulation}. 
The simulated results, depicted in~\fig{fig: CMDMresults}, reveal a high-pass nature in the frequency response curve of CM current(~\fig{fig: CMDMresults}(a)). Furthermore,~\fig{fig: CMDMresults}(b) and~\fig{fig: CMDMresults}(c) illustrate the attacker's ability to optimize the frequency of the attack signal, thereby enhancing the CM-DM conversion efficiency. In summary, we draw the following conclusions: (1) The injected attack signal can couple to the signal line parallel to the victim device's signal GND, subsequently generating a CM current that is then converted to a DM voltage at the output load. (2) Both the coupling and conversion efficiency are influenced by the asymmetric impedance and the signal frequency.


\subsection{Asymmetric Impedance in Electronic Circuits} \label{sec: asymmetric impedance}
In this section, we explore and summarize three key factors that affect asymmetric impedance, which plays a critical role in energy conversion as discussed in~\mysec{sec: energy_conversion_model}, and validate our analysis with simulations and experiments.



\textbf{Asymmetric design in electronic circuits.} Asymmetric impedance can arise from variations in the quantities and electronic characteristics of fundamental electronic components (e.g., resistors, capacitors, etc.) across different circuits. For instance, designers may use different combinations of resistance and capacitance to achieve specific time constants and frequency responses in RC circuits or select different transistors to supply adequate driving currents for loads. 
Additionally, the arrangement of electronic components on a PCB impacts the equivalent impedance, especially for high-frequency signals~\cite{1997Designing}. 
This occurs because the parasitic capacitance and inductance between components vary with frequency~\cite{ADIPCBDesign,carter2009circuit}, altering the parasitic impedance~\cite{Anand2019StudyonPCB}. 
Despite the introduction of various optimization strategies, achieving perfectly symmetric designs remains a challenge.

\textbf{Inconsistent parameters caused by manufacturing.} Imperfect manufacturing processes can result in inaccuracies in the values and physical characteristics of electronic components. Variations in raw materials, environmental conditions during production, and unavoidable deviations in the manufacturing process contribute to inconsistencies. Furthermore, parameter shifts and performance degradation occur over time due to aging effects. 
In practice, mass-produced electronic components typically adhere to a tolerance range around the standard value after delivery. For example, a resistor rated at 1$\Omega$ may measure 1.001$\Omega$. 
Another common case involve amplifiers, which are widely used in electronic circuits to amplify DM signals and suppress CM signals. 
However, due to the inconsistency in transistor characteristics~\cite{analog2009op}, such as gain and threshold voltage, achieving complete CM signal elimination is unattainable.
This limitation is quantified by the common-mode rejection ratio (CMRR)~\cite{analog2009op}: $CMRR_i \approx 2g_m^2R_{ss}/\Delta g_m $, where $g_m$ denotes the average transconductance of two input transistors, $\Delta g_m$ represents the difference in transconductance between the transistors ($\Delta g_m=g_{m1}-g_{m2}$), and $R_{ss}$ is the finite output impedance of current source in the amplifier~\cite{razavi2005design}. The formula illustrates that greater asymmetry in the input transistors lead to a weaker CM rejection capability, highlighting the impact of transistor inconsistencies on amplifier performance. 



\begin{figure} [!t]  
	\centering
	\subfigure[Setup for op-amp.]{
		\includegraphics[height=0.34\linewidth]{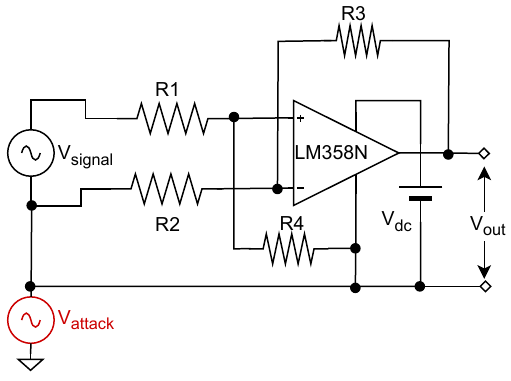}
		\label{fig: AMPAttack}
	}\hspace{-2ex}
        \subfigure[Setup for RS flip-flop.]{
		\includegraphics[height=0.34\linewidth]{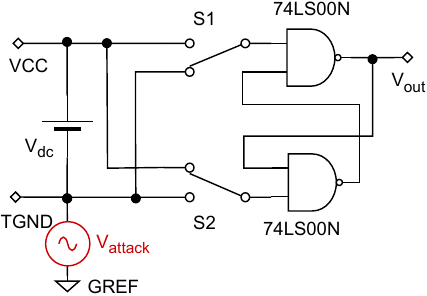}
		\label{fig: RSAttack}
	}
	\subfigure[Output Signal of op-amp.]{
		\hspace{-2ex}
            \includegraphics[height=0.19\linewidth]{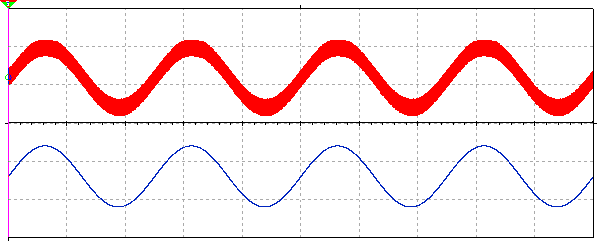}
		\label{fig: AMPAttackResult}
	}
        \subfigure[Output Signal of RS flip-flop.]{
		\includegraphics[height=0.19\linewidth]{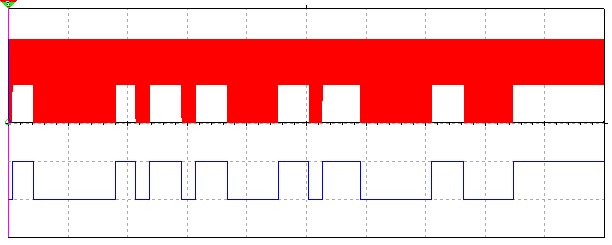}
		\label{fig: RSAttackResult}
	}
	\caption{Illustration of simulation of (a) an op-amp and (b) an RS flip-flop. When injecting a crafted attack signal into the GND, (c)the amplifier's output(blue) will be superimposed by a sinusoidal signal(red) due to imperfect CMRR, and (d) the RS flip-flop's original output(blue) will be induced with many false pulses(red) due to asynchronous operations.}
 \vspace{-0.3cm}
\end{figure}

\begin{figure}[!t]  
    	\centering
    	\subfigure[Output of Amplifier (LM386).]{
    		\includegraphics[width=0.45\linewidth]{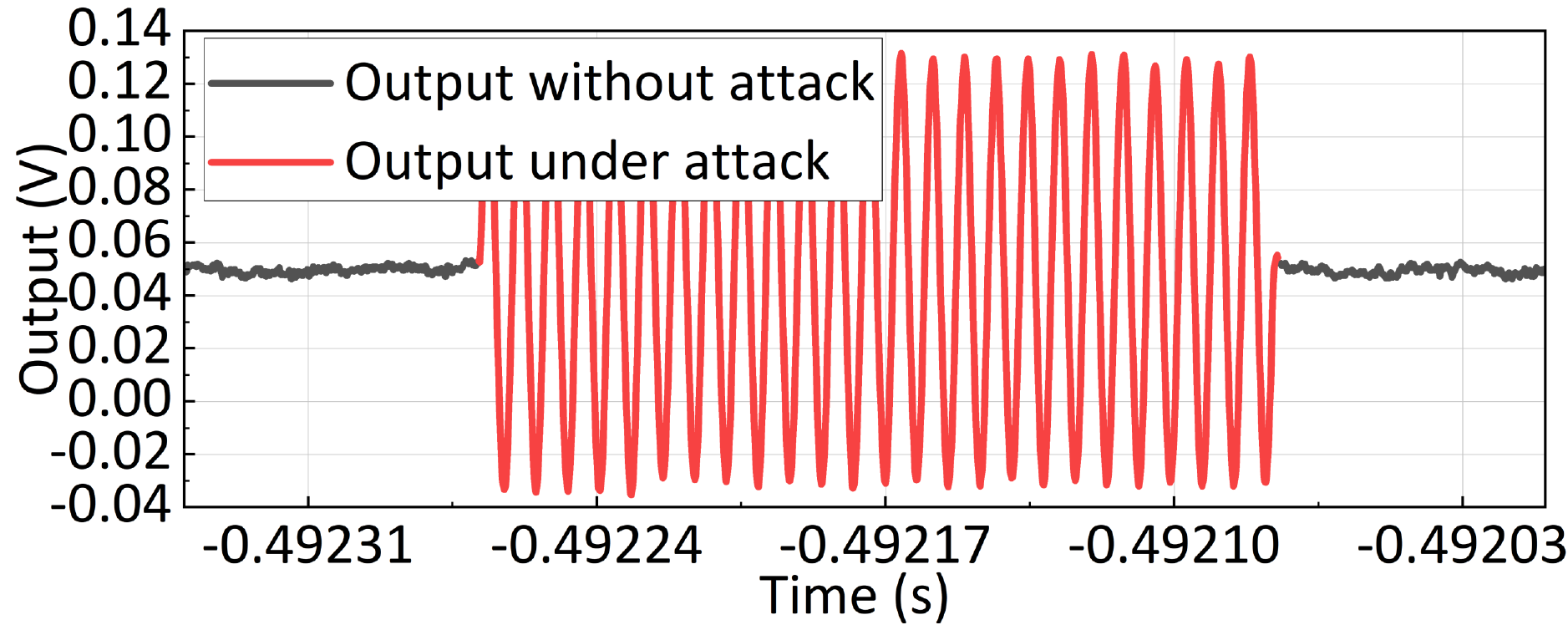}
    		\label{fig: amp_injection}
    	}
     \hspace{-2ex}
    	\subfigure[Output of VFC (LM331).]{
    		\includegraphics[width=0.45\linewidth]{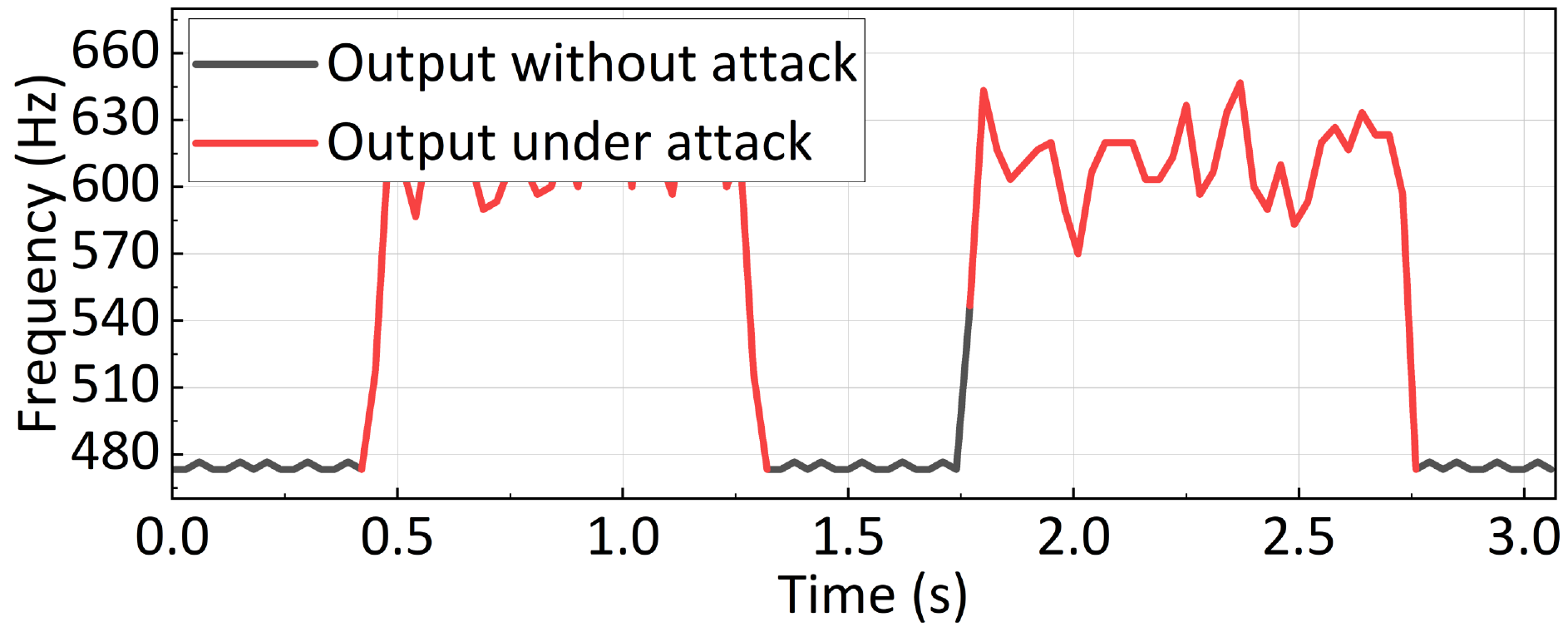}
    		\label{fig: vfc_injection}
    	}
    	\caption{\blue{ Experiments results of an amplifier and a VFC. (a)~\alias introduced a sinusoidal signal to the amplifier's (LM386) output. (b)~\alias modified the output of a VFC (LM331) from 473Hz to 643Hz.}}
    	\label{fig: physical_experiment}
    \end{figure}

\textit{Validation.} To verify our analysis, we conducted validation experiments. (a) Simulation: We used the Multisim tool~\cite{multisim} to construct a typical differential amplifier circuit with an operational amplifier (LM358N) as shown in~\fig{fig: AMPAttack}, where $V_{attack}$ represents the attack voltage source. The output waveforms of the amplifier with (red waveform) and without attacks (blue waveform) are presented in~\fig{fig: AMPAttackResult}. These results demonstrate the efficient traversal of a CM signal through the differential amplifier, resulting in the injection of a false AC signal into the output. (b) Experiments: Additionally, physical experiments were conducted using an LM386 amplifier development board, driven by an Arduino Uno. A crafted CM signal (introduced in~\mysec{sec: design}) was injected into the GND port. The results are shown in ~\fig{fig: amp_injection}, where the black waveform represents the normal output of the amplifier without an input signal, and the red waveform indicates the malicious output triggered by the attack. In conclusion, both simulation and physical experiments confirm that CM interference significantly affects the amplifier's output.

\textbf{Asynchronous action of dynamic devices.} 
Another asymmetry factor is the asynchronous behavior of dynamic electronic components, such as MOSFETs and IGBTs. 
These switching components are commonly used to regulate and manage current flows by turning circuits on and off. 
Take the RS flip-flop as an example, where two switching components are aligned across two transmission lines to operate simultaneously. The circuit's impedance changes as the internal conduction state of the switching elements transitions. 
However, maintaining synchronization between two switches is challenging in practical applications due to clock skew, which arises from differences in the clock signal's arrival times across various segments~\cite{hauck1995asynchronous}. 
Furthermore, state mismatches between controllers and switching elements~\cite{shi2019asynchronous}, caused by delays in the transmission of the control signals and the switching operation, represent another contributing factor to asynchronous behavior.


\textit{Validation.} We performed validation experiments on an RS flip-flop, a widely used component in digital electronics, such as memory elements or latches. (a) Simulation: The simulation circuit, shown in~\fig{fig: RSAttack}, consists of two NAND gates latched to each other. The normal output of the flip-flop is displayed as the blue waveform in~\fig{fig: RSAttackResult}, with levels alternate between high and low. After injecting a malicious CM signal into the GND, the flip-flop's output oscillates frequently between high and low levels, demonstrating the attack's impact. (b) Experiments: Physical experiments were conducted on a voltage-to-frequency converter (VFC) utilizing the LM331, with the RS flip-flop as its critical component. The results, shown in~\fig{fig: vfc_injection}, indicate that the presence of the CM signal (311.15kHz, 300Vpp) can alter the output frequency, causing a shift from 473Hz to 643Hz, further validating our analysis.

%% file: sections/Design.tex
\section{Attack Design}
\label{sec: design}
To achieve~\alias, we face two technical challenges: (1) How to effectively inject attack signals into the target sensor via the power cable? (2) How to create a given output? In this section, we present the attack design of~\alias.

\subsection{Effective Signal Injection}\label{sec: effective_injection}
To effectively inject attack signals into the target sensor via the power cable and enhance the interference effect, we first select available signals by optimizing attack parameters.

\begin{figure} [!t]  
	\centering
	\subfigure[Frequency response.]{
		\includegraphics[width=0.47\linewidth, height = 0.25\linewidth]{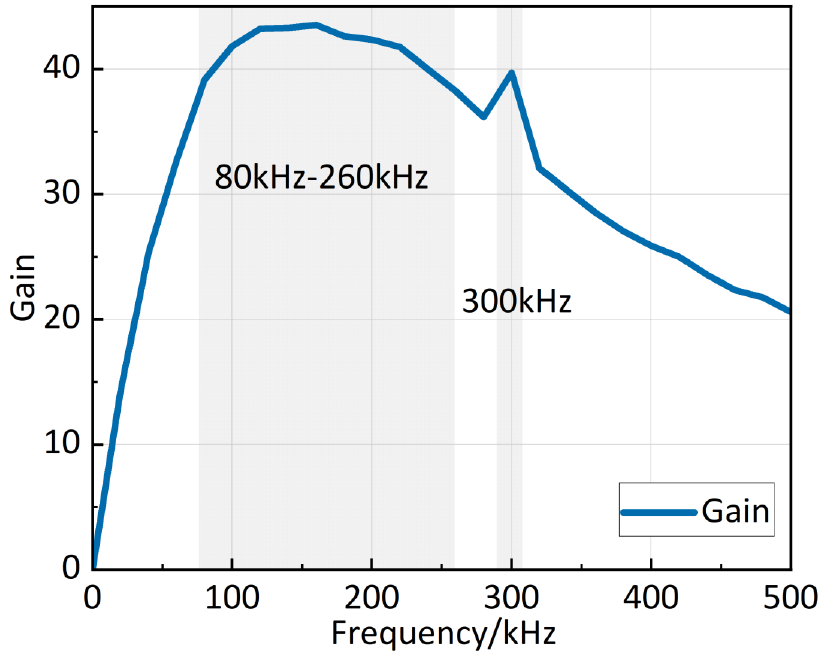}
		\label{fig: AMP_gain_f}
	}\hspace{-0.4em}
	\subfigure[Amplitude response.]{
		\includegraphics[width=0.47\linewidth, height = 0.25\linewidth]{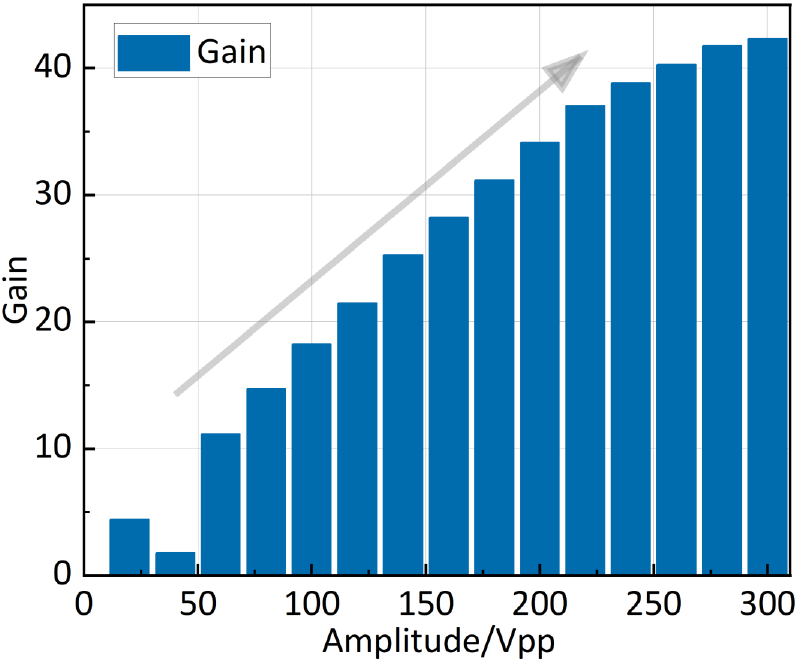}
		\label{fig: AMP_gain_v}
	}\hspace{-0.4em}
	\caption{Frequency response and amplitude response of an amplifier (LM386) under~\alias attack. (a) When the CM signal is in  [80kHz, 260kHz] and 300kHz, the frequency response of the amplifier is relatively high. (b) The result shows a stronger attack signal generally can induce a higher output.}
\end{figure}

\textbf{Signal frequency optimization.} As described in the energy conversion model introduced in~\mysec{sec: feasibility}, both the coupling and conversion coefficient vary with the signal frequency change. 
Given the complexity of electric circuits and components in IoT devices, we employ a frequency sweep strategy to evaluate overall performance from signal injection to final output. 
Specifically, we sweep the attack signal and identify frequencies that result in relatively large false outputs or severe performance degradation as vulnerable frequencies for the victim device.
Additionally, we conducted a preliminary experiment on an op-amp development board (LM386) (the experiment setup is introduced in~\mysec{sec: evaluation}). A signal with a voltage of 300\,V and frequencies ranging from 0\,Hz to 500\,kHz was injected into the GND line of the LM386. The frequency response, shown in~\fig{fig: AMP_gain_f}, suggests that the 80\,kHz to 260\,kHz and 300\,kHz frequency ranges are particularly suitable as attack frequencies due to their large amplification gain deviation of over 38.

\textbf{Signal amplitude optimization.} Similar to the signal frequency, the amplitude of the attack signal plays a crucial role in energy conversion. As illustrated in \eq{eq: ICM_Vs}) and \eq{con: define_vdmo} in~\mysec{sec: feasibility}, a proportional relationship exists between the DM output voltage and the intensity of the CM signal, making the impact of signal amplitude evident. To validate this relationship, We conducted an experimental study on an LM386 amplifier, varying the attack signal amplitude between 0\,V to 300\,V, and recorded the responses at intervals of 20\,V.~\fig{fig: AMP_gain_v} demonstrates that a stronger signal generally results in more significant interference with the victim device.

\subsection{Attack Signal Design} \label{sec: attack_signal_design}
Although attackers can effectively induce random interference to the sensor's output by optimizing attack parameters, their ultimate goal is to create a given output. Therefore, the attacker should \blue{design} more fine-grained attack signals.

 
	

\begin{figure} [!t]  
	\centering

	\subfigure[Nonlinearity-based AC injection method.]{
		\includegraphics[width=0.29\linewidth]{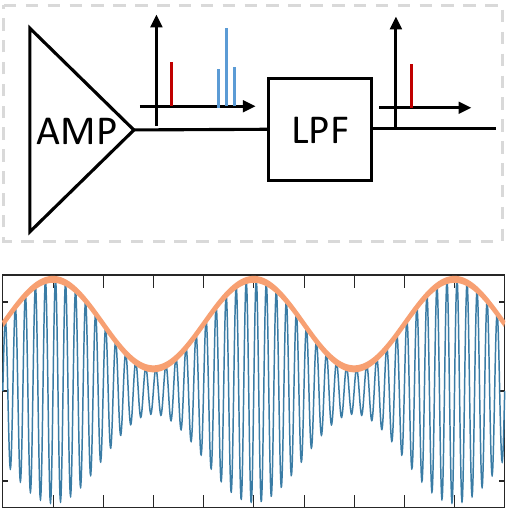}
		\label{fig: ac}
	}
	\subfigure[Jitter-based pulse injection method.]{
		\includegraphics[width=0.29\linewidth]{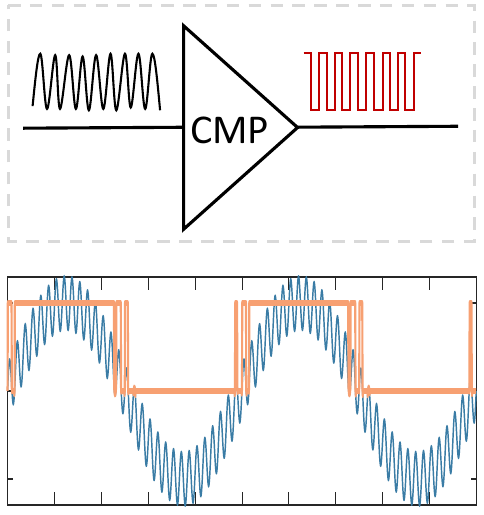}
		\label{fig: pulse}
	}
 	\subfigure[Biasing-based DC injection method.]{
		\includegraphics[width=0.29\linewidth]{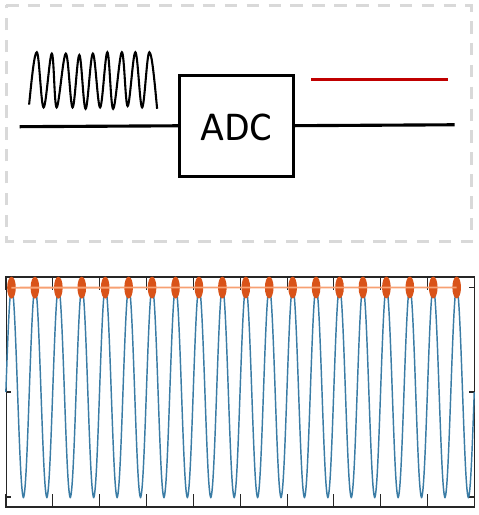}
		\label{fig: dc}
	}
	\caption{Illustrations of three attack signal design methods. The attacker can achieve stealthy injection by leveraging the inherent imperfections of electronic components and crafting attack signals to emulate legitimate outputs, such as AC, pulses, or DC signals. }
 \vspace{-1.0em}
 \label{fig: attacksignal_design}
	
\end{figure}

\textbf{Baseband signal design.} An attacker first prepares the baseband signal, which is typically associated with the target output signal and is intended to be demodulated by the target sensor. For instance, if the attacker aims to inject an inaudible voice command into a microphone, the target voice command signal serves as the baseband signal.

\textbf{Manipulating signal crafting.} Since the attacker has limited control over the shape of the injected signal, they can only exploit the vulnerabilities inherent in the target sensor's circuits to fine-tune its output. We present three signal crafting methods as follows: 

\textit{(a) Modulation-based AC injection method.} This method exploits the nonlinearity vulnerability of electronic components, such as amplifiers~\cite{zhang2017dolphinattack}, to generate new frequency AC signals. By utilizing frequency selection circuits (e.g., low-pass filters (LPF)) to demodulate desired signals, this approach ensures specific signal components are retained. An illustration of the signal design is shown in~\fig{fig: ac}. Let the input signal be denoted as $s_{in}(t)$, then the output signal $s_{out}$ can be expressed as~\cite{zhang2017dolphinattack}: $s_{out} = A s_{in}(t) + B s^2_{in}(t)$, where $A$ is the gain for the input signal, and $B$ is the gain for the quadratic term $s^2_{in}$. Suppose the desired output of the victim device is $m(t)=\cos(2 \pi f_m t)$, where $f_m$ is the frequency and the vulnerable frequency of the device is $f_c$. The attack signal can then be chosen as $s_{in}(t) = (m(t)+1) \cos (2 \pi f_c t)$. After applying $s_{in}(t)$ to the output equation, taking the Fourier transform, and processing it through a frequency selection circuit, the desired component $f_m$ can remain in the output. 

\underline{\textit{Applicable scene:}} This method is suitable for sensors such as microphones that require the injection of low-frequency AC signals and incorporate nonlinear components like amplifiers, which can effectively demodulate desired signals from the attack signal. For example, if the vulnerable frequency of the microphone module EG8542 is 370kHz and the desired voice command $m(t)$, the attack signal can be designed as $s(t) = (m(t)+1) \cos (2\times 370 \times 1000 \pi t)$.

\textit{(b) Jitter-based pulse injection method.} This method exploits the over-sensitivity vulnerability of electronic components such as comparators, which are highly sensitive to input noise or interference. Pulse-output devices, such as speed sensors and rotary encoders, typically utilize a combination of an amplifier and a comparator~\cite{analog2014speedsensor}. The amplifier boosts the weak output signal of the transducer, while the hysteresis comparator converts the amplified signal into a digital output. This cascade structure is inherently vulnerable, as it combines the vulnerabilities of both the amplifier and the comparator, making it susceptible to false pulse generation. Specifically, an attacker can select a vulnerable signal frequency, as described in~\mysec{sec: effective_injection}, to introduce jitter at the amplifier's output, as illustrated by the blue waveform in~\fig{fig: pulse}. When the jitter causes the output to exceed the threshold of the hysteresis comparator, the output signal is pulled up, otherwise, it is pulled down, resulting in fake pulses. 

\blue{\underline{\textit{Applicable scene:}} This method is suitable for pulse-based output sensors, such as speed sensors, rotary encoders, and motion detection sensors. These sensors typically include a combination of an amplifier and a comparator and are designed to process pulse signals. For instance, an attacker can inject pulses into a rotary encoder to spoof the speed measurements, as described in~\mysec{sec: attack_on_sensors}.}

\textit{(c) Biasing-based DC injection method.} The third type of signal design scheme is the biasing-based DC injection method, in which the attacker leverages the sampling distortion vulnerability of the ADC to reshape fluctuating false measurements into a stabilized bias. Assuming the ADC's sampling frequency is $f_s$ Hz, and the signal frequency is $f_N$, according to the Nyquist-Shannon sampling theorem, $ f_s $ must be at least twice $ f_N $ to avoid aliasing. Otherwise, aliases will be produced, and the aliased frequency $ f_a $ of the reconstructed signal can be expressed as: $ f_a = | 2mf_N - f_s | $, where $ m $ is an integer such that $f_a < f_N$. In this case, the attacker can manipulate the frequency of the output signal. For example, if $f_a=0$, the output signal becomes a stabilized DC bias, as illustrated in~\fig{fig: dc}. Similar to the signal modulation strategy in~\cite{trippel2017walnut}, an attacker can also utilize amplitude modulation (AM), phase modulation (PM), or frequency modulation (FM) to craft the output signal arbitrarily. 

\blue{\underline{\textit{Applicable scene:}} This method is suitable for sensors where the output signal is a constant value, such as distance sensors, accelerometers, gyroscopes. It is particularly effective when the ADC's sampling rate is lower than the carrier signal, as a low sampling rate enables the attacker to create frequency aliasing, resulting in a new frequency. By applying this method, the victim sensor can be manipulated to output a stable and falsified value.}

%% file: sections/Evaluation.tex
\section{\ Implementation and Evaluation}
\label{sec: evaluation}
We introduce the implementation and overall performance.

\begin{figure}[!t]  
	\centering
	\includegraphics[width=1\linewidth]{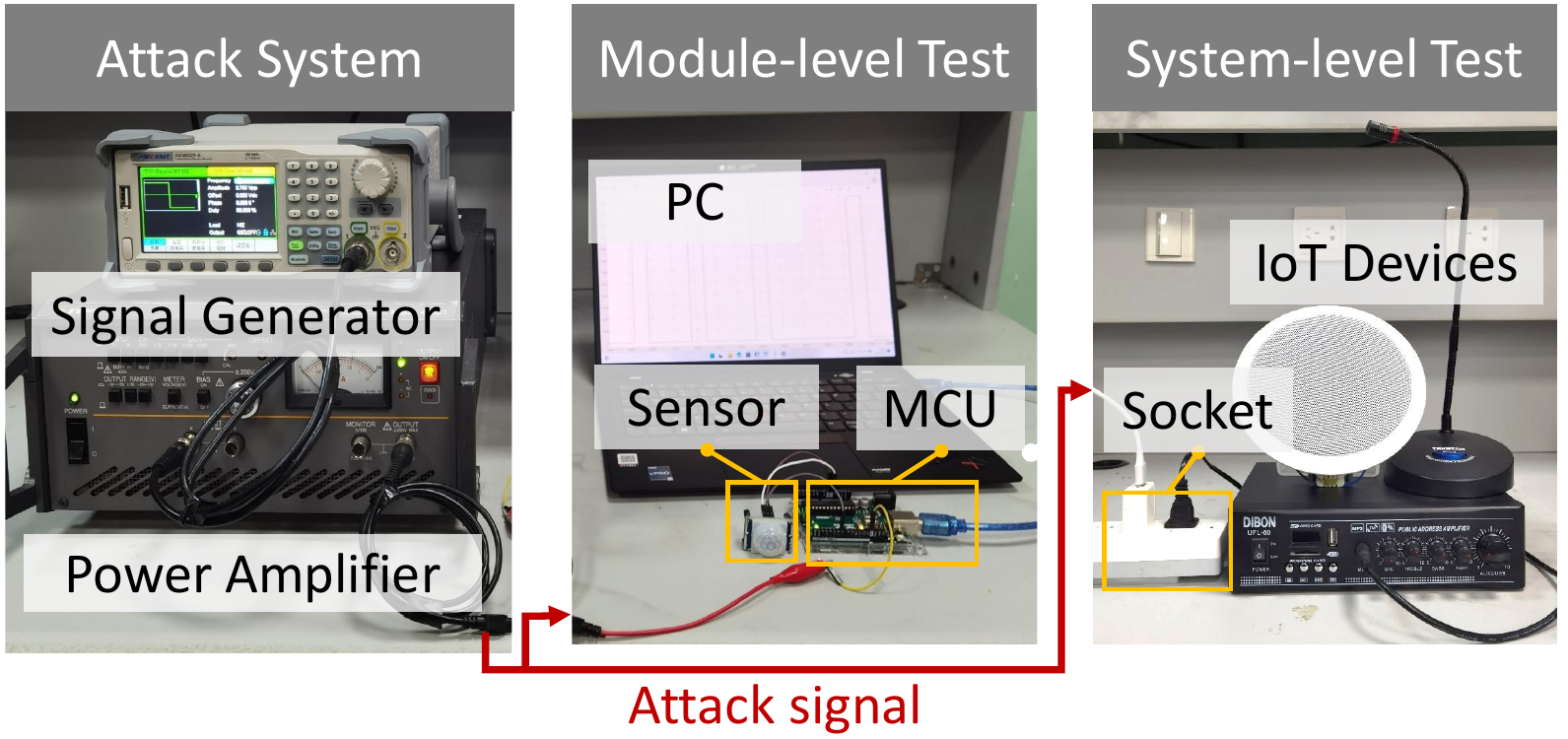}
	\caption{Evaluation setup of module-level (including electronic modules and sensors) and system-level, where the attack system consists of a signal generator and a power amplifier. A malicious signal is injected into the GND wire of the tested device via the driver board or a charging socket.}
	\label{fig: setup_sensor}
\end{figure}

\input{tables/table_evaluation_functionalmodule}

\subsection{Experimental Setup}\label{sec: setup}
\textbf{Attack System.} The attack system consists of an arbitrary waveform generator (SDG6032X~\cite{SDG6032X}) and a power amplifier (NFHSA4051~\cite{2013NFHSA4051} for the low-frequency range and ZHL100WGAN~\cite{ZHL100WGAN} for the high-frequency range), as depicted in~\fig{fig: setup_sensor}. The output of the signal generator is connected to the input port of the power amplifier~\cite{2013NFHSA4051, ZHL100WGAN}. The positive output of the amplifier is connected to the signal GND of the target device, while the negative output of the amplifier is left floating. For ethical considerations, we don't inject interference into active power grids. Instead, we use a grid testbed to replicate the wired environment for experiments. Specifically, an uninterruptible power supply (UPS) (BULL GNV-Y16~\cite{bullUPS}) serves as the power source (e.g., 220V and 50Hz), and power strips are used to mimic an electronic distribution network connecting various electric loads. 

\textbf{Test Objects.} To evaluate the effectiveness and universality of~\alias, we selected target devices spanning various categories, design solutions, and brands. The test objects were classified into the following three categories: \textit{(a) Electronic Modules.} We evaluated 13 signal-processing modules across 6 categories: Amplifiers (AMP), Voltage-to-Frequency converters (VFC), Frequency-to-Voltage converters (FVC), Analog-to-Digital converters, Digital-to-Analog converters (DAC), signal generations (GEN). Detailed models are provided in~\tab{tab: functionalmodules}. \textit{(b) Sensors.} We tested 17 common sensors across 9 categories: light sensors, microphones, encoders, vibration sensors, distance sensors, water detection sensors, motion sensors, accelerators, and hall sensors. \textit{(c) Real-world Systems.} We applied~\alias to a surveillance system and a broadcast system. To validate the practicality of~\alias, we conducted cross-socket attacks, as shown in the right panel of~\fig{fig: setup_sensor}, by injecting attack signals into the GND port of the socket.




\textbf{Signal Parameters and Driver Board.} We selected signal parameters that maximize attack performance while remaining within the device's capabilities (300Vpp, 500kHz). Additionally, when evaluating~\alias on functional modules and sensors, we used a microcontroller (MCU) (Arduino UNO~\cite{arduino2023arduinouno}) to drive these modules and transmit the attack signal to the GND port of the MCU, as shown in the middle figure of~\fig{fig: setup_sensor}.

\subsection{Overall Performance}
In this section, we present the overall performance of~\alias. The metrics used to evaluate the attack capability include the success rate and output deviation.
\begin{itemize}[leftmargin=5mm]
    \item \textit{Success rate:} The proportion of attacks that successfully induce false outputs in the tested objects.
    \item \textit{Output deviation:} The difference between the false output and the original value, which quantifies the attack's impact on the output accuracy of the tested device.
\end{itemize}

\subsubsection{Attack on Electronic Modules} 
We evaluated 13 functional electronic modules across six categories using the setup shown in~\fig{fig: setup_sensor} (middle), where the MCU drives the modules and the PC displays their output. The detailed attack parameters are provided in~\tab{tab: functionalmodules}. 
\textit{(a) AMPs:} AMPs are widely used in IoT devices to increase the signal-to-noise ratio (SNR) of measurement signals. However, due to their non-ideal properties, amplifiers may amplify CM noises. Experimental results demonstrate that~\alias successfully induces false AC signals on all tested AMP modules' outputs. For example, with an attack signal of 80\,kHz and 300\,Vpp,~\alias alters the AD623's output from 2.5\,V to 3.89\,V. Since the LM386 has a lower common-mode rejection ratio than the THS3091, it exhibits a severe negative impact, resulting in a deviation of 98.7\,\%. Imagine the AMP is applied to critical scenarios such as the precision machining industry, incorrect outputs could lead to significant errors. 
\textit{(b) Data converters:} Data converters play a vital role in linking the analog and digital domains by converting data between these forms. For instance, an ADC digitizes an analog wave for processing, while a DAC performs the reverse, converting a digital code into an analog signal. In this study, we evaluated four common ADCs and one DAC. The results demonstrate that~\alias can disrupt the conversion process, inducing errors. For instance,~\alias causes the digital output of ADS 1100 from 5395 to 13 and shifts the DAC902's output voltage from 0.01\,V to 0.08\,V. In industrial applications such as machine tools, even a small ADC error could result in severe consequences, such as a machine crash.
\textit{(c) Other functional modules:} We also evaluateD other functional modules, including VFCs, FVCs, and signal generators. A VFC produces an output signal frequency proportional to its control voltage, while a signal generator can be seen as a specialized VFC. Conversely, an FVC converts the frequency of an input signal into a proportional output voltage. The results in~\tab{tab: functionalmodules} confirm the feasibility of~\alias. For example,~\alias shifts the output frequency of the VFC (LM331) from 467Hz to 700Hz. Furthermore, we observed that the attacker can control the output by adjusting the amplitude of the attack signal. For instance, reducing the attack signal strength lowers the output frequency of the VFC NE555.


\begin{figure} [!t]  
	\centering
        \subfigure[Experimental setup.]{
		\includegraphics[height=0.38\linewidth]{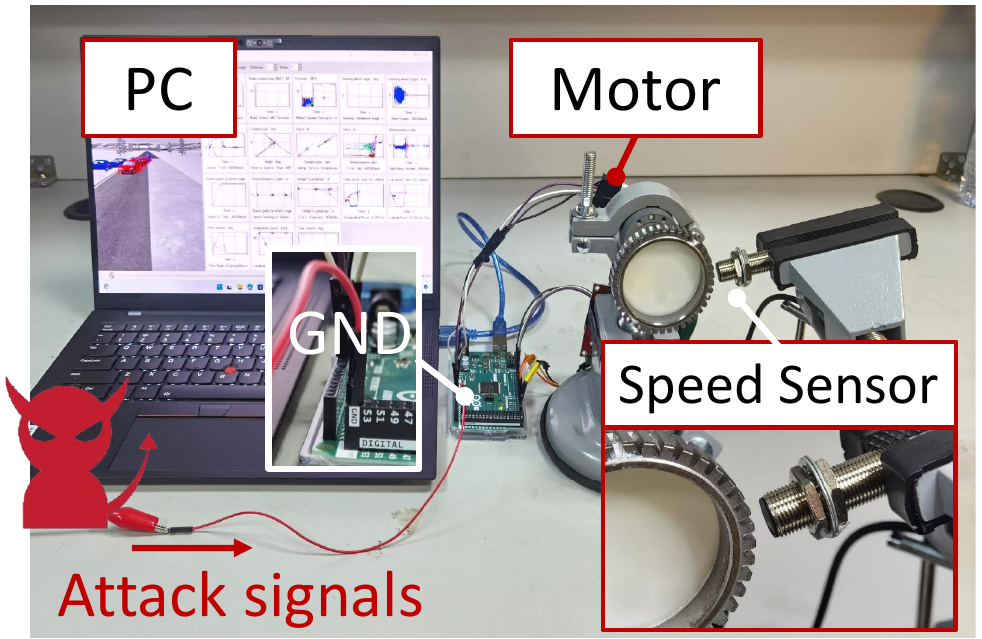}\label{fig: setup_car}
	}\hspace{-1ex}
        \subfigure[Simulation results.]{
		\includegraphics[height=0.38\linewidth]{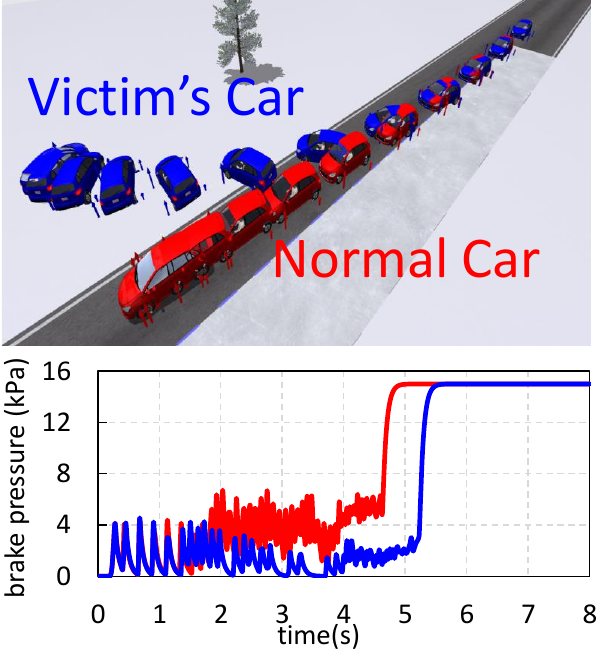}\label{fig: results_car}
	}
        \caption{Evaluation of a speed sensor (encoder) with a hardware-in-the-loop (HIL) autonomous driving simulation system.}
 \label{fig: speedsensor_evaluation}
  \vspace{-0.3cm}
\end{figure}

\input{tables/table_evaluation_sensormodule}

\subsubsection{Attack on Sensors} \label{sec: attack_on_sensors}
We tested 17 widely used sensors, and the results illustrated in~\tab{tab: sensormodules} demonstrate that sensors are susceptible to~\alias. For example,~\alias can induce a sinusoidal signal on the output of the microphone sensor \blue{(EG8542), leading to an increase in} the peak-to-peak output magnitude from 13 to 155. \blue{We also observe that some microphones exhibit greater resilience to~\alias due to their high CMRR performance and short analog signal wires, such as the TDA1308.} Additionally,~\alias \blue{can also} spoof the security system, leading to a false alarm by targeting the HCSR05 distance sensor. Due to space limitation, detailed attack parameters for other sensor modules are provided in~\tab{tab: sensormodules} without individual introductions. 
Furthermore, we developed a Hardware-In-the-Loop (HIL) autonomous driving simulation using the CarSim ~\cite{CarSim2023} and MATLAB Simulink tool~\cite{HIL2023mathworks}. This system facilitates testing of autonomous driving systems by receiving speed measurements from a commercial ABS sensor. The experimental setup is depicted in~\fig{fig: setup_car}, where the PC runs the simulation software, providing calculated speed data to the physical motor (i.e. the speed baseline), and the Arduino Mega drives the ABS sensor while communicating with the PC. During the attack, we injected a 250\,kHz, 300\,Vpp attack signal into the GND pin of the Arduino Mega. The results, shown in the right figures of~\fig{fig: results_car}, highlight that~\alias can effectively manipulate the speed sensor, causing the victim's car (blue car) to lose control of its speed and deviate from its predetermined route. In a real-world attack scenario, an attacker could attach a modular attack device to the vehicle's chassis and connect the attack signal output to the vehicle chassis to access the vehicle's GND~\cite{cadence2024effective}. A similar threat model, involving attaching the attack device to the chassis, has been demonstrated in prior research~\cite{shoukry2013non}.

\begin{figure}[!t]  
    \centering
    \includegraphics[width=0.95\linewidth]{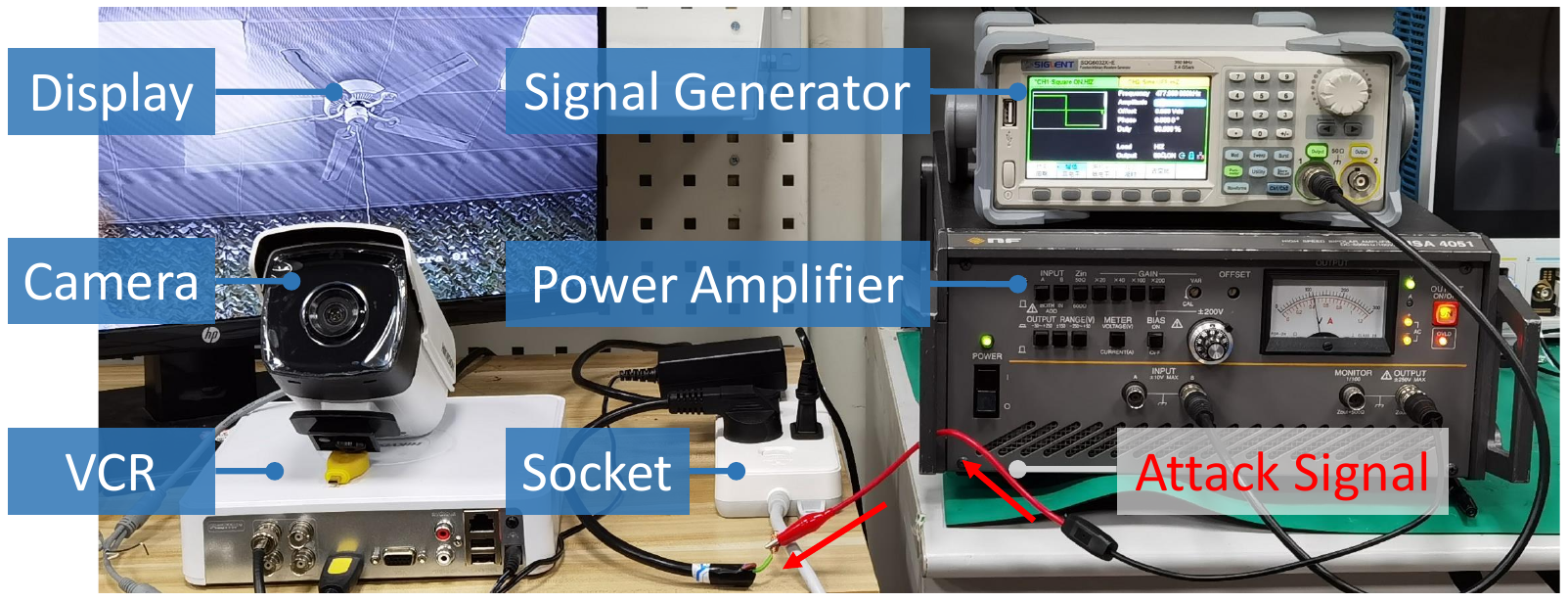}
	\caption{Evaluation on a surveillance system. We conduct a cross-socket attack against a surveillance system by introducing stripes \blue{into the captured images}.}
	\label{fig: setup_camera}
\end{figure}

\begin{figure} [!t]  
	\centering
        \subfigure[Facenet Results.]{
		\includegraphics[height=0.37\linewidth]{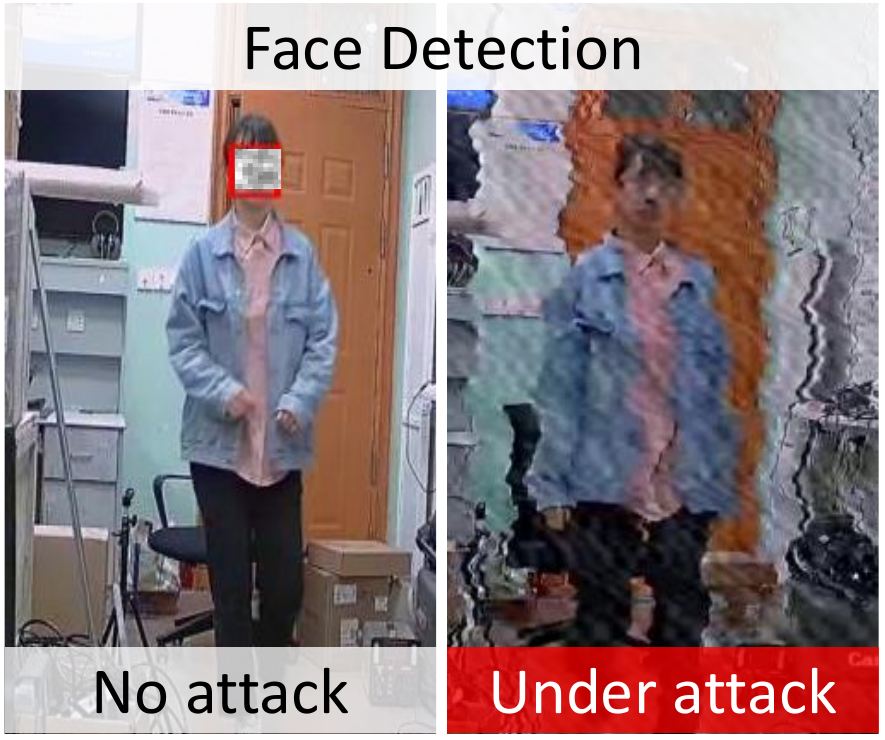}\label{fig: eval_facenet}
	}\hspace{2ex}
        \subfigure[Yolov8 Results.]{
		\includegraphics[height=0.37\linewidth]{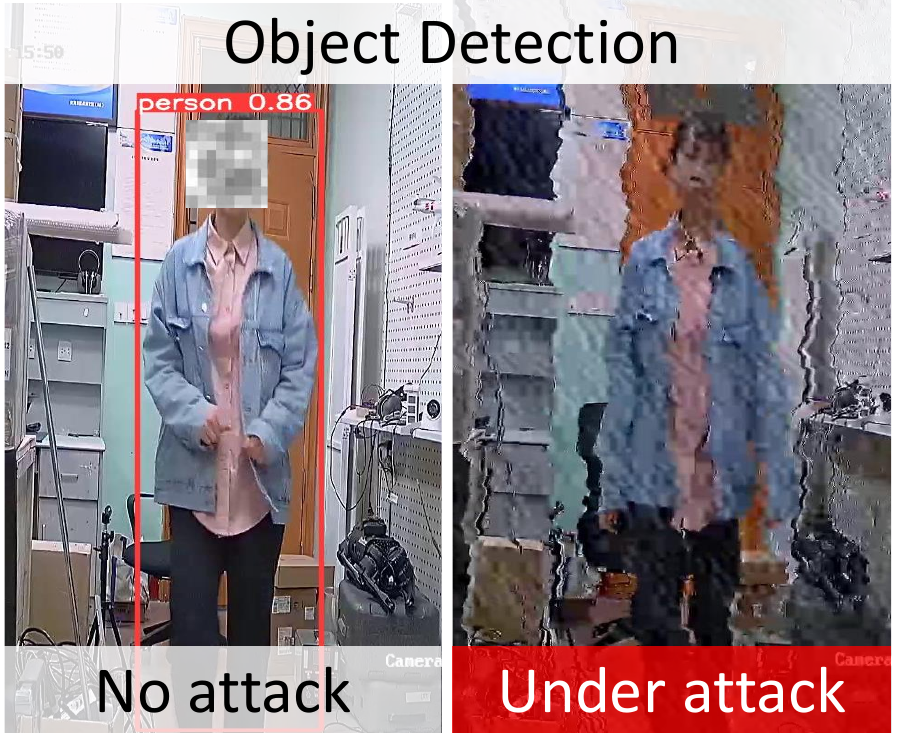}\label{fig: eval_yolov8}
	}
	\caption{Results of face detection and object detection in real-world attacks. When conducting~\alias, (a) the face detection model (Facenet~\cite{schroff2015facenet}) fails to detect the intruder's face, and (b) the object detection model (Yolov8~\cite{jocher2023yolov8}) cannot identify the intruder.}
 \label{fig: eval_face_object}
\end{figure}

\subsubsection{Attack on Real-world Systems} \label{sec: real-world attack}
To demonstrate the real-world threat posed by~\alias, we focus on attacking two critical systems: surveillance systems and broadcast systems, by manipulating cameras and microphones. 
\textbf{Surveillance System.} Surveillance systems are widely deployed in smart homes, airports, etc. Analog cameras, in particular, are popular in such systems due to their cost-effectiveness and robustness compared to digital cameras~\cite{datainteloXXXXglobal}. 
In our study, we uncover a new attack vector for analog cameras, where an attacker can induce stripes on the images captured by the camera. 
The underlying mechanism involves the analog signal wire of the camera coupling with the injected signal on the GND wire, generating a common-mode (CM) current as described in~\mysec{sec: feasibility}. Due to the asymmetric structure within the camera, this CM current is converted into an alternating DM voltage, which interferes with analog transmission signals such as video and sync signals, resulting in stripes on the captured images.~\fig{fig: setup_camera} illustrates a home surveillance system attack scenario, where the video cassette recorder (VCR) receives the analog signal generated by the commercial high-definition (HD) camera (HIKVISION DS-2CE56C3T-IT3~\cite{hikvisiono2024ds2ce16d0t}) and transmits the video signals to the display. By injecting attack signals (477.9\,kHz, 170\,Vpp) into the GND wire of the camera via a shared socket, stripes are introduced into the captured images. To highlight the real-world implications of this attack, we evaluate~\alias on a face detection model (Facenet~\cite{schroff2015facenet}) and an object detection model (Yolov8~\cite{jocher2023yolov8}). The captured images, along with the detection results, are shown in~\fig{fig: eval_face_object}, illustrating that~\alias can effectively manipulate the analog HD camera and disrupt surveillance capabilities. 


\input{tables/table_evaluation_voice_complexity}

\textbf{Broadcast System.} Broadcast systems are pivotal for information dissemination in public spaces such as airports and supermarkets, playing a crucial role in ensuring public safety and social stability. We evaluated~\alias on a broadcast system by injecting inaudible voice commands into a commercial broadcast microphone (TAKSTAR MS-118~\cite{takstar2024ms118}). A typical broadcast system consists of a desktop microphone, an audio modulator, and a speaker, as shown in~\fig{fig: setup_sensor}. Specifically, the microphone is connected to the audio modulator (UFL-60~\cite{mrslm2024ufl}) for voice audio amplification, and the modulator's output is connected to a speaker. 
Similar to the attack on the surveillance system, we injected attack signals into the GND wire of the shared socket. 
The attacker first selects a vulnerable attack signal, as introduced in~\mysec{sec: design}. Next, malicious voice audio, such as ``Attention, please!" is prepared and modulated onto the vulnerable signal using the modulation-based AC injection method. Due to the frequency selection circuits in the microphone, such as low-pass filtering circuits, the malicious audio is demodulated from the attack signal and subsequently played through the speaker. Appendix~\fig{fig: results_audio} shows the waveforms and spectrograms of the recorded audios: (a) the target audio signal $S_t$ and (b) the injected audio $S_i$ both played by the broadcast system. 

To evaluate the quality of the injected audio, we employed Wav2Vec~\cite{baevski2020wav2vec}, a commonly used algorithm for extracting speech content features, to calculate the cosine similarity between the injected audio $S_i$ and the target audio $S_t$. To account for audio distortion caused by the broadcast system, the similarity between the raw audio $S_r$ and the target audio $S_t$ is used as the baseline (BL).
Additionally, to verify the audibility and intelligibility of the injected audio, we utilize an automatic speech recognition (ASR) model, Distil-Whisper~\cite{gandhi2023distilwhisper} to transcribe the recordings to text sequences and calculate the Levenstein distance~\cite{levenshtein1966binary} (L-dis), a measurement of the similarity between two strings.
Furthermore, to demonstrate the versatility of~\alias in voice injection scenarios, we evaluate 8 common voice commands that are widely used in broadcasting scenarios, e.g., the flight broadcast, and fire alarm, with varying complexity. The detailed voice commands and results are listed in~\tab{tab: voice complexity}. The $W2V$ results show the similarities closely approximate the baseline, around 0.6. Despite slight distortion in the injected audio due to energy attenuation of low-frequency signals during the signal coupling stage, the findings suggest that~\alias can effectively inject voice commands. Furthermore, the L-dis results confirm that~\alias can successfully inject voice audio with accurate semantics into the microphone, spoofing the broadcast system with diverse phrases.



\subsubsection{\blue{Home Wiring Scenario.}}\label{sec: homegrid}
\begin{figure} [!t]  
	\centering
	\subfigure[Home wiring scenario.]{
		\includegraphics[height=0.5\linewidth]{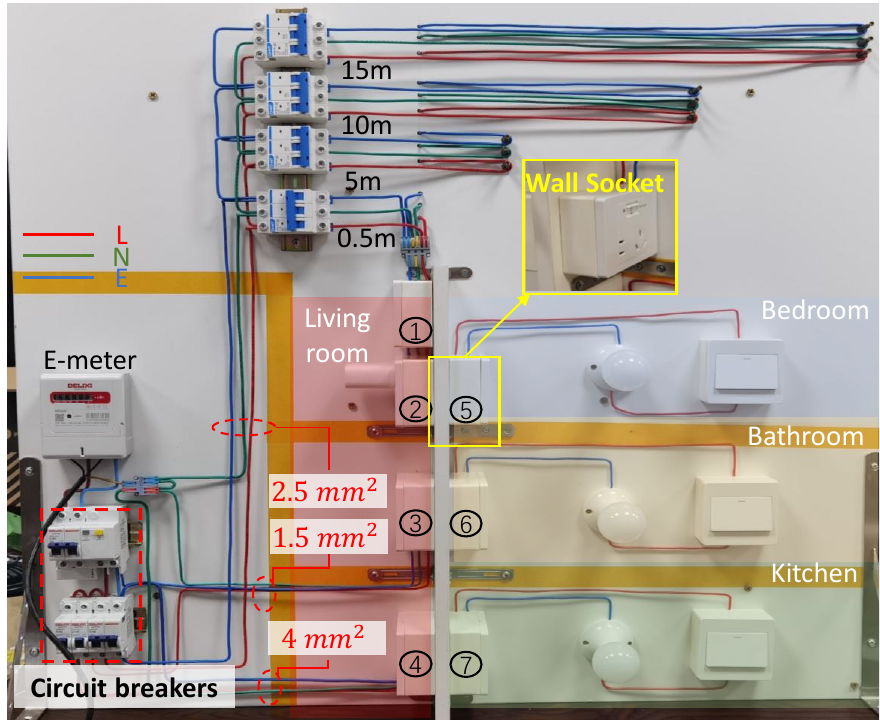}
		\label{fig: microgrid_1}
	}
	\subfigure[Attack setup.]{
		\includegraphics[height=0.5\linewidth]{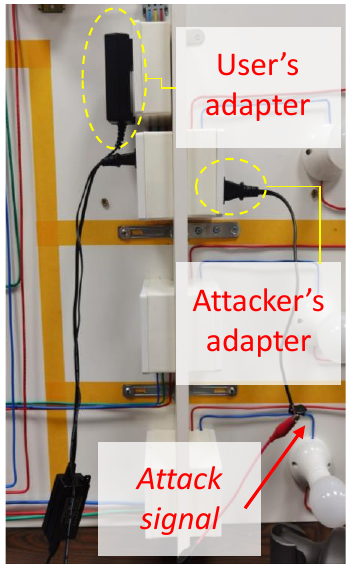}
		\label{fig: microgrid_2}
	}
    \caption{\blue{(a) Illustration of a home wiring scenario, which encompasses a power source, power distribution, and power consumption subsystem. (b) Illustration of the real-world attack setup.}}.
 \label{fig: microgrid}	
 \vspace{-1em}
\end{figure}

To verify the effectiveness of~\alias in real-world wired environments,such as home scenarios, we established a household power system, as shown in~\fig{fig: microgrid}. This system comprises a power source (UPS), power distribution systems, and power consumption systems. In this section, we evaluate four factors that may influence the attack's effectiveness by conducting~\alias on this system.

\input{tables/table_evaluation_microgrid}

\textbf{Setup.} The attack setup is shown in the upper-right figure of~\fig{fig: microgrid}, where the attack signal (320\,kHz, 260Vpp) is injected into the attacker's adapter, plugged into a wall socket, while the broadcast system is plugged into another wall socket. We conducted~\alias on a tested broadcast system (TAKSTAR MS-118) and evaluated the signal similarities and semantics of injected audios (``Attention, Please'') as described in~\ref{sec: attack_on_sensors}. The evaluation factors and results are detailed below.

\blue{\textbf{Factors and Results.} 
\textbf{(1) Circuit breaker:} The household power system shown in~\fig{fig: microgrid} includes four circuit breakers~\cite{scherz2006practical} (1-pole and 2-pole types) designed to protect electrical circuits from overcurrent or short circuits. Since~\alias relies on CM voltage rather than current, circuit breakers do not block the attack signal. A physical experiment validated this analysis, yielding Wav2Vec (W2V) similarity scores of 0.69 and Levenshtein distance (L-dis) of 0, demonstrating that~\alias remains effective in a home wiring scenario with circuit breakers.
\textbf{(2) Wiring types:} We tested three specifications with cross-sectional areas of 1.5\, $mm^2$, 2.5\, $mm^2$, and 4\, $mm^2$, supporting maximum currents of 16\,A, 25\,A and 36\,A, respectively. Each test used one wiring type, and the results demonstrate~\alias is effective across all wiring specifications.
\textbf{(3) Electrical system layout:} To assess the impact of electrical system layout on~\alias, we evaluated three practical attack scenarios, in-room (e.g., socket~\ding{172} and~\ding{173}), cross-wall (e.g., socket~\ding{173} and~\ding{176}) and cross-room (e.g., socket~\ding{173} and~\ding{178}). Results depicted in~\tab{tab: eval_microgrid} confirm the effectiveness of~\alias under all tested configurations.
\textbf{(4) Electrical noise:} We simulated various noise conditions in a household power system using seven IoT devices, including a desktop, fan, smart speaker, smartphone, and three bulbs. Experimental results show that~\alias successfully injects voice commands into the broadcast system under different noise conditions.}
\input{tables/table_evaluation_camera}
\subsection{Other Factors}
We discuss and evaluate potential factors that may affect the performance of~\alias, \blue{including the device models, the driver board and the attack distance.}

\subsubsection{Device Models.} To access the transferability of~\alias, we evaluated its performance on various device models.

\textbf{Cameras:} We evaluated~\alias on eight off-and-shelf HD surveillance cameras from five international brands (HIKVISION~\cite{hikvisiono2024ds2ce16d0t,hikvisiono2024ds2ce56cot}, Dahua~\cite{dahua2024dhhac}, Panasonic~\cite{panasonic2024wvcw314lCH}, SAMSUNG~\cite{samsung2024sco2080r}, and SONY~\cite{sony2024imx323}), as shown in Appendix~\ref{sec: supplement_eval}~\fig{fig: eval_camera_list}. The evaluation setup is shown in~\fig{fig: setup_camera} and the detailed attack parameters are provided in~\tab{tab: eval_camera}. During the experiments, the cameras recorded an experimenter entering the room while~\alias was active. Captured frames or videos were analyzed using two detectors (YOLOv8~\cite{jocher2023yolov8}, Facenet~\cite{schroff2015facenet}). Baseline detection results were obtained from unaltered images, and the attack was deemed successful if the detectors failed to identify objects or faces.
The success rate of this attack is defined as the ratio between the number of misidentified frames to the total number of frames. Attack results in~\tab{tab: eval_camera} indicate that all tested cameras were vulnerable to~\alias. A 100\% attack success rate was achieved for six out of eight cameras. Notably, the Panasonic camera's adaptive digital noise reduction~\cite{panasonic2024wvcw314lCH} made it more challenging to inject deep stripes compared to other models. However, lighter stripes were sufficient to evade detection systems.

\textbf{Microphones:} We also evaluated~\alias against seven commercial desktop microphones (HUAWEI AM115~\cite{productz2024huaweimic}, HP DHP-1100l~\cite{jd2024hpmic}, Lenovo Lecoo MC01 ~\cite{banggood2024lenovmic}, UGREEN CM564~\cite{jg2024ugreen}, SM88, TAKSTAR MS-118~\cite{takstar2024ms118}, HIKVISION DS-KAU30HG-M~\cite{jd2024ds-kau30hg-m}), covering seven brands and three interface types (3.5\,mm jack, USB port, XLR connector), as shown in Appendix~\ref{sec: supplement_eval}~\fig{fig: eval_mic_list}. Detailed attack parameters are given in~\tab{tab: eval_mic}, where the auxiliary devices are used to play the fake injected audios. 
Similar to~\mysec{sec: real-world attack}, we evaluated the quality of the injected voice signals using Levenstein distance~\cite{levenshtein1966binary}. During the experiments. While most microphones exhibited effective voice injection, we observed reduced clarity in the UGREEN CM564, likely due to its unique circuit structure and shorter analog signal lines. Results in~\tab{tab: eval_mic} confirm that~\alias can successfully inject inaudible voice commands into all tested microphones, further demonstrating its versatility.

\input{tables/table_evaluation_commercialmic}
\subsubsection{\blue{Effect on the Driver Board}}
In the experiments, we used microcontrollers such as the Arduino UNO~\cite{arduino2023arduinouno} and STM32~\cite{stm2020stm32} to drive the tested electronic modules and sensors, and a laptop to display the output data.
To ensure that~\alias successfully injects malicious signals into the sensor's analog signal wire rather than interfering with the microcontroller's power supply or communication, we employed a TIVP02 differential high-voltage probe~\cite{TIVP022020tektronix}. The probe offers high CMRR performance and accessibility, allowing precise measurements of the analog output of the tested module, the power supply of the driver board, and the digital communication between the driver board and the laptop. 
The experimental setup is shown in Appendix~\fig{fig: tivp02_setup}, where the frequency of the attack signal is set to 300\,kHz. The results are shown in~\fig{fig: dm_probe_test}. The top figure of~\fig{fig: dm_probe_test} shows the analog output of the tested module, demonstrating that~\alias successfully injects malicious AC signals. The middle and bottom figures of~\fig{fig: dm_probe_test} validate our analysis that the power supply and digital communication remain unaffected by the attack signal.

\subsubsection{Attack Distance}
Additionally, we assess the impact of attack distance on~\alias. Specifically, we constructed four alternative electrical wiring paths with distances of 0.5\,m, 5\,m, 10\,m, 15\,m as shown in~\fig{fig: microgrid}. These paths can be switched using control switches, and the additional wires are wound on the back panel, as shown in Appendix A~\fig{fig: back panel}. We evaluated cross-wall attacks on a microphone (TAKSTAR MS-118) and a camera (DS-2CE56D8T-IT3) using attack signal at 320\,kHz, 260\,Vpp and 478\,kHz, 300\,Vpp, respectively. The results in~\tab{tab: attack distance} demonstrate the feasibility of~\alias across various attack distances. The essential reason for this effectiveness is that~\alias exploits the power cable to transmit CM voltage instead of DM current, resulting in minimal energy loss along the electrical wiring. Compared to similar wireless EMI research focused on tasks such as manipulating sensor measurements~\cite{shoukry2013non}, injecting voice commands~\cite{kasmi2015iemi}, or interfering with image sensors~\cite{kohler2022signal},~\alias achieves a longer attack distance of up to 15\,m. For example, \cite{kasmi2015iemi} was capable of injecting inaudible voice commands at a maximum distance of 4\,m, using 200\,W of power, while~\cite{kasmi2015iemi} interfered with image sensors at a maximum distance of 50\,cm.

\input{tables/table_evaluation_attack_distance}


%% file: tables/table_evaluation_functionalmodule.tex



\begin{table*}[]
\footnotesize 
\centering  
\caption{Evaluation of~\alias attacks on 13 functional modules of 6 categories: Amplifiers (AMP), Voltage-to-Frequency converters (VFC), Frequency-to-Voltage converters (FVC), Analog-to-Digital converters, Digital-to-Analog converters (DAC), signal generations (GEN).} %
\label{tab: functionalmodules}
\setlength\tabcolsep{6pt} 
\renewcommand{\arraystretch}{1.3} 

\begin{tabular}{|p{0.7cm}<{\centering}|p{1.2cm}<{\centering}|p{0.5cm}<{\centering}p{0.5cm}<{\centering}|p{0.6cm}<{\centering}p{0.6cm}<{\centering}p{0.9cm}<{\centering}|p{0.7cm}<{\centering}|p{1.5cm}<{\centering}|p{0.5cm}<{\centering}p{0.5cm}<{\centering}|p{0.6cm}<{\centering}p{0.6cm}<{\centering}p{0.9cm}<{\centering}|}
\hline
\multirow{2}{*}{\textbf{Type}} & \multirow{2}{*}{\textbf{Model}} & \multicolumn{2}{c|}{\textbf{Parameters}}           & \multicolumn{3}{c|}{\textbf{Output}}                                                        & \multirow{2}{*}{\textbf{Type}} & \multirow{2}{*}{\textbf{Model}} & \multicolumn{2}{c|}{\textbf{Parameters}}                         & \multicolumn{3}{c|}{\textbf{Output}}                                                                              \\ \cline{3-7} \cline{10-14} 
                                  &                                 & \multicolumn{1}{c|}{\textbf{fre.}} & \textbf{vpp.} & \multicolumn{1}{c|}{\textbf{org.}} & \multicolumn{1}{c|}{\textbf{att.}} & \textbf{dev.(\%)} &                                   &                                 & \multicolumn{1}{c|}{\textbf{fre.}}        & \textbf{vpp.}        & \multicolumn{1}{c|}{\textbf{org.}}         & \multicolumn{1}{c|}{\textbf{att.}}         & \textbf{dev.(\%)}       \\ \hline
\multirow{4}{*}{\textbf{AMP}}     & AD623                           & \multicolumn{1}{c|}{80}            & 300           & \multicolumn{1}{c|}{2.5}           & \multicolumn{1}{c|}{3.89}          & 55.6             & \multirow{4}{*}{\textbf{ADC}}     & ADS1100                         & \multicolumn{1}{c|}{120M}                 & 0.3                  & \multicolumn{1}{c|}{5395}                  & \multicolumn{1}{c|}{13}                    & 99.7                   \\ \cline{2-7} \cline{9-14} 
                                  & AD620                           & \multicolumn{1}{c|}{160}           & 300           & \multicolumn{1}{c|}{3.85}          & \multicolumn{1}{c|}{4.7}           & 22.0             &                                   & AD7606                          & \multicolumn{1}{c|}{190M}                 & 1                    & \multicolumn{1}{c|}{0}                     & \multicolumn{1}{c|}{1567}                  & 100.0                  \\ \cline{2-7} \cline{9-14} 
                                  & THS3091                         & \multicolumn{1}{c|}{500}           & 300           & \multicolumn{1}{c|}{1.37}          & \multicolumn{1}{c|}{1.41}          & 2.9              &                                   & MCP4725                         & \multicolumn{1}{c|}{500}                  & 300                  & \multicolumn{1}{c|}{254}                   & \multicolumn{1}{c|}{271}                   & 6.6                    \\ \cline{2-7} \cline{9-14} 
                                  & LM386                           & \multicolumn{1}{c|}{130}           & 300           & \multicolumn{1}{c|}{398}           & \multicolumn{1}{c|}{791}           & 98.7             &                                   & STM32F103                       & \multicolumn{1}{c|}{500}                  & 300                  & \multicolumn{1}{c|}{1.23}                  & \multicolumn{1}{c|}{1.32}                  & 7.3                    \\ \hline
\multirow{2}{*}{\textbf{VFC}}     & LM331                           & \multicolumn{1}{c|}{311.15}        & 260           & \multicolumn{1}{c|}{467Hz}         & \multicolumn{1}{c|}{700Hz}         & 49.9             & \multirow{2}{*}{\textbf{DAC}}     & \multirow{2}{*}{DAC902}         & \multicolumn{1}{c|}{\multirow{2}{*}{500}} & \multirow{2}{*}{300} & \multicolumn{1}{c|}{\multirow{2}{*}{0.01}} & \multicolumn{1}{c|}{\multirow{2}{*}{0.08}} & \multirow{2}{*}{700.0} \\ \cline{2-7}
                                  & NE555                           & \multicolumn{1}{c|}{380}           & 300           & \multicolumn{1}{c|}{210Hz}         & \multicolumn{1}{c|}{2100Hz}        & 900.0            &                                   &                                 & \multicolumn{1}{c|}{}                     &                      & \multicolumn{1}{c|}{}                      & \multicolumn{1}{c|}{}                      &                         \\ \hline
\textbf{FVC}                      & LM331                           & \multicolumn{1}{c|}{500}           & 300           & \multicolumn{1}{c|}{3.69V}         & \multicolumn{1}{c|}{3.72V}         & 0.8              & \textbf{GEN}                      & NE555                           & \multicolumn{1}{c|}{470}                  & 260                  & \multicolumn{1}{c|}{467Hz}                 & \multicolumn{1}{c|}{1133Hz}                & 142.6                  \\ \hline
\end{tabular}
\begin{tablenotes}[para,flushleft]
\item \textbf{Note:} \emph{fre.} and \emph{vpp} indicate the frequency (kHz) and the voltage amplitude (V) of the attack signal respectively. \emph{org.} represents the original output of the tested modules before conducting attacks, and \emph{att.} indicates the false output under~\alias attacks. Additionally, \emph{dev.} represents the deviation ratio, calculated as the percentage ratio of the difference between the measured value and the actual value to the actual value itself.
\end{tablenotes}
 \vspace{-0.3cm}
\end{table*}

%% file: tables/table_evaluation_sensormodule.tex
\begin{table*}[]

\footnotesize 
\centering  
\caption{We evaluated 17 common sensors of 9 categories including light sensors (light), microphones (mic.), encoders, vibration sensors, distance sensors, water detection sensors, motion sensors, accelerators (acc.), and pedal sensors (pedal).} 
\label{tab: sensormodules}
\setlength\tabcolsep{6pt} 
\renewcommand{\arraystretch}{1.2} 


\begin{tabular}{|p{1cm}<{\centering}|p{1.5cm}<{\centering}|p{0.5cm}<{\centering}p{0.5cm}<{\centering}|p{0.6cm}<{\centering}p{0.6cm}<{\centering}p{0.9cm}<{\centering}|p{1.2cm}<{\centering}|p{1.5cm}<{\centering}|p{0.5cm}<{\centering}p{0.5cm}<{\centering}|p{0.6cm}<{\centering}p{0.6cm}<{\centering}p{0.9cm}<{\centering}|}
\hline
\multirow{2}{*}{\textbf{Sensors}} & \multirow{2}{*}{\textbf{Model}} & \multicolumn{2}{c|}{\textbf{Parameters}}                         & \multicolumn{3}{c|}{\textbf{Output}}                                                                       & \multirow{2}{*}{\textbf{Sensors}} & \multirow{2}{*}{\textbf{Model}} & \multicolumn{2}{c|}{\textbf{Parameters}}           & \multicolumn{3}{c|}{\textbf{Output}}                                                        \\ \cline{3-7} \cline{10-14} 
                                  &                                 & \multicolumn{1}{c|}{\textbf{fre.}}        & \textbf{vpp.}        & \multicolumn{1}{c|}{\textbf{org.}}       & \multicolumn{1}{c|}{\textbf{att.}}       & \textbf{dev.(\%)}    &                                   &                                 & \multicolumn{1}{c|}{\textbf{fre.}} & \textbf{vpp.} & \multicolumn{1}{c|}{\textbf{org.}} & \multicolumn{1}{c|}{\textbf{att.}} & \textbf{dev.(\%)} \\ \hline
\textbf{light}                    & CGMCU101                        & \multicolumn{1}{c|}{120}                  & 180                  & \multicolumn{1}{c|}{929}                 & \multicolumn{1}{c|}{1024}                & 10.2                 & \multirow{3}{*}{\textbf{encoder}} & E6B2                            & \multicolumn{1}{c|}{380}           & 300           & \multicolumn{1}{c|}{2066}          & \multicolumn{1}{c|}{15000}         & 626.0            \\ \cline{1-7} \cline{9-14} 
\multirow{8}{*}{\textbf{mic.}}    & LM386                           & \multicolumn{1}{c|}{100}                  & 300                  & \multicolumn{1}{c|}{0.371}               & \multicolumn{1}{c|}{1.37}                & 269.3                &                                   & GMR                             & \multicolumn{1}{c|}{350}           & 210           & \multicolumn{1}{c|}{129}           & \multicolumn{1}{c|}{660}           & 411.6            \\ \cline{2-7} \cline{9-14} 
                                  & MAX4466                         & \multicolumn{1}{c|}{0.15}                 & 300                  & \multicolumn{1}{c|}{0.17}                & \multicolumn{1}{c|}{0.59}                & 247.1                &                                   & ABS                             & \multicolumn{1}{c|}{250}           & 300           & \multicolumn{1}{c|}{10}            & \multicolumn{1}{c|}{320}           & 3100.0           \\ \cline{2-14} 
                                  & MK519                           & \multicolumn{1}{c|}{500}                  & 300                  & \multicolumn{1}{c|}{0.01}                & \multicolumn{1}{c|}{0.78}                & 7700.0               & \textbf{vibration}                    & SW18010P                        & \multicolumn{1}{c|}{27}            & 300           & \multicolumn{1}{c|}{0}             & \multicolumn{1}{c|}{1}             & 100.0            \\ \cline{2-14} 
                                  & \multirow{2}{*}{TDA1308}        & \multicolumn{1}{c|}{\multirow{2}{*}{150}} & \multirow{2}{*}{280} & \multicolumn{1}{c|}{\multirow{2}{*}{53}} & \multicolumn{1}{c|}{\multirow{2}{*}{58}} & \multirow{2}{*}{9.4} & \textbf{distance}                    & HCSR04                          & \multicolumn{1}{c|}{306}           & 300           & \multicolumn{1}{c|}{153}           & \multicolumn{1}{c|}{0}             & 100.0            \\ \cline{8-14} 
                                  &                                 & \multicolumn{1}{c|}{}                     &                      & \multicolumn{1}{c|}{}                    & \multicolumn{1}{c|}{}                    &                      & \textbf{water}                    & LM393                           & \multicolumn{1}{c|}{410}           & 300           & \multicolumn{1}{c|}{1.87}          & \multicolumn{1}{c|}{2.12}          & 13.3            \\ \cline{2-14} 
                                  & EG8542                          & \multicolumn{1}{c|}{370}                  & 300                  & \multicolumn{1}{c|}{13}                  & \multicolumn{1}{c|}{155}                 & 1092.3               & \textbf{motion}                   & HCSR05                          & \multicolumn{1}{c|}{90}            & 300           & \multicolumn{1}{c|}{0}             & \multicolumn{1}{c|}{1}             & 100.0            \\ \cline{2-14} 
                                  & CJMCU622                        & \multicolumn{1}{c|}{170}                  & 300                  & \multicolumn{1}{c|}{500}                 & \multicolumn{1}{c|}{690}                 & 38.0                 & \textbf{acc.}                     & ADXL345                         & \multicolumn{1}{c|}{49.93}         & 280           & \multicolumn{1}{c|}{0g}            & \multicolumn{1}{c|}{2g}            & 100.0            \\ \cline{2-14} 
                                  & MAX9814                         & \multicolumn{1}{c|}{220}                  & 230                  & \multicolumn{1}{c|}{235}                 & \multicolumn{1}{c|}{280}                 & 19.1                 & \textbf{pedal}                    & Hall                            & \multicolumn{1}{c|}{160}           & 300           & \multicolumn{1}{c|}{2°}            & \multicolumn{1}{c|}{40°}           & 1900.0          \\ \hline
\end{tabular}
 \vspace{-0.3cm}
\end{table*}

%% file: tables/table_evaluation_voice_complexity.tex
\begin{table}[t]
\footnotesize 
\centering  
\caption{\blue{Evaluation of voice complexity}} 
\label{tab: voice complexity}
\setlength\tabcolsep{3.5pt} 
\renewcommand{\arraystretch}{1.2} 
\begin{tabular}{cccccc}
\toprule
\# & \textbf{Voice Commands}              & \textbf{Scenarios}   & \textbf{BL} & \textbf{W2V}  & \textbf{L-dis} \\ \hline
1  & ``Keep your phone switched off''     & Airport       & 0.73 & 0.63  & 0     \\
2  & ``Flight will arrive at platform''   & Airport       & 0.84 & 0.60  & 13     \\
3  & ``Attention, please''                & Fire alarm    & 0.79 & 0.69  & 0     \\
4  & ``Fire alarm activated''             & Fire alarm    & 0.82 & 0.54  & 7      \\
5  & ``Please evacuate the building''     & Market        & 0.83 & 0.65  & 0     \\
6  & ``Deadline is approaching''          & Office        & 0.82 & 0.62  & 0    \\
7  & ``Stay indoors''                     & Weather       & 0.85 & 0.63  & 0     \\
8  & ``Tomorrow will have showers''       & Weather       & 0.50 & 0.58  & 2     \\
\bottomrule
\end{tabular}
\vspace{-2em}
\end{table}

%% file: tables/table_evaluation_microgrid.tex
\begin{table}[]
\footnotesize 
\centering  
\caption{\blue{Results of 4 types of factors in a household power system. The results demonstrate the effectiveness of~\alias in practical attack scenarios.}} 
\label{tab: eval_microgrid}
\setlength\tabcolsep{2.2pt} 
\renewcommand{\arraystretch}{1.2} 
\begin{tabular}{|c|c|ccccccc|}
\hline
\multirow{2}{*}{\textbf{Factors}} &
  \multirow{2}{*}{\textbf{BL}} &
  \multicolumn{1}{c|}{\multirow{2}{*}{\textbf{\begin{tabular}[c]{@{}c@{}}Circuit \\      Breakers\end{tabular}}}} &
  \multicolumn{3}{c|}{\textbf{Wiring   Types}} &
  \multicolumn{3}{c|}{\textbf{Electrical   System Layouts}} \\ \cline{4-9} 
 &
   &
  \multicolumn{1}{c|}{} &
  \multicolumn{1}{c|}{1.5} &
  \multicolumn{1}{c|}{2.5} &
  \multicolumn{1}{c|}{4} &
  \multicolumn{1}{c|}{in-room} &
  \multicolumn{1}{c|}{cross-wall} &
  cross-room \\ \hline
\textbf{W2V} &
  0.79 &
  \multicolumn{1}{c|}{0.69} &
  \multicolumn{1}{c|}{0.75} &
  \multicolumn{1}{c|}{0.60} &
  \multicolumn{1}{c|}{0.72} &
  \multicolumn{1}{c|}{0.55} &
  \multicolumn{1}{c|}{0.55} &
  0.54 \\ \hline
\textbf{L-dis} &
  0 &
  \multicolumn{1}{c|}{0} &
  \multicolumn{1}{c|}{0} &
  \multicolumn{1}{c|}{0} &
  \multicolumn{1}{c|}{0} &
  \multicolumn{1}{c|}{0} &
  \multicolumn{1}{c|}{0} &
  0 \\ \specialrule{1pt}{0pt}{0pt}
\multirow{2}{*}{\textbf{Factors}} &
  \multirow{2}{*}{\textbf{BL}} &
  \multicolumn{7}{c|}{\textbf{Electrical Noises}} \\ \cline{3-9} 
 &
   &
  \multicolumn{1}{c|}{desktop} &
  \multicolumn{1}{c|}{fan} &
  \multicolumn{2}{c|}{speaker} &
  \multicolumn{1}{c|}{phone} &
  \multicolumn{1}{c|}{1 bulb} &
  3 bulbs \\ \hline
\textbf{W2V} &
  0.79 &
  \multicolumn{1}{c|}{0.55} &
  \multicolumn{1}{c|}{0.54} &
  \multicolumn{2}{c|}{0.55} &
  \multicolumn{1}{c|}{0.54} &
  \multicolumn{1}{c|}{0.52} &
  0.54 \\ \hline
\textbf{L-dis} &
  0 &
  \multicolumn{1}{c|}{0} &
  \multicolumn{1}{c|}{0} &
  \multicolumn{2}{c|}{0} &
  \multicolumn{1}{c|}{0} &
  \multicolumn{1}{c|}{0} &
  0 \\ \hline
\end{tabular}
\vspace{-1em}
\end{table}

%% file: tables/table_evaluation_camera.tex
\begin{table}[]
\footnotesize 
\centering  
\caption{Evaluation on 8 camera models. All of these commercial cameras are vulnerable to~\alias \blue{and the captured images can avoid being detected by the object detector (Yolov8) and the face detector (Facenet).}} 
\label{tab: eval_camera}
\setlength\tabcolsep{3pt} 
\renewcommand{\arraystretch}{1.2} 

\begin{tabular}{|c|cc|c|cc|}
\hline
\multirow{2}{*}{Camera   Model} & \multicolumn{2}{c|}{  Parameters} & \multirow{2}{*}{\begin{tabular}[c]{@{}c@{}}Inject\\ Stripe\end{tabular}} & \multicolumn{2}{c|}{Success   rate}    \\ \cline{2-3} \cline{5-6} 
                                & \multicolumn{1}{c|}{fre.}      & amp.    &                                                                          & \multicolumn{1}{c|}{Facenet} & Yolov8  \\ \hline
HIKVISION DS-2CE56D8T-IT3       & \multicolumn{1}{c|}{478}       & 140     & \Checkmark                                                                      & \multicolumn{1}{c|}{99.0\%}  & 100.0\% \\ \hline
HIKVISION DS-2CE16G0T-IT3       & \multicolumn{1}{c|}{477.9}     & 170     & \Checkmark                                                                      & \multicolumn{1}{c|}{98.5\%}  & 100.0\% \\ \hline
DH-HAC-HFW1200M-I2              & \multicolumn{1}{c|}{450.2}     & 150     & \Checkmark                                                                      & \multicolumn{1}{c|}{89.1\%}  & 100.0\% \\ \hline
Panasonic WV-CW314LCH           & \multicolumn{1}{c|}{485.3}     & 270     & \Checkmark                                                                      & \multicolumn{1}{c|}{59.7\%}  & 75.5\%  \\ \hline
SAMSUNG SCO-2080RP              & \multicolumn{1}{c|}{411.2}     & 310     & \Checkmark                                                                      & \multicolumn{1}{c|}{88.2\%}  & 100.0\% \\ \hline
SONY CCD673-1200                & \multicolumn{1}{c|}{468.6}     & 200     & \Checkmark                                                                      & \multicolumn{1}{c|}{89.7\%}  & 93.1\%  \\ \hline
SONY CCD-1200                   & \multicolumn{1}{c|}{453}       & 200     & \Checkmark                                                                      & \multicolumn{1}{c|}{98.6\%}  & 100.0\% \\ \hline
SONY IMX323                     & \multicolumn{1}{c|}{506.2}     & 120     & \Checkmark                                                                      & \multicolumn{1}{c|}{97.4\%}  & 100.0\% \\ \hline
\end{tabular}
\vspace{-0.5cm}
\end{table}

%% file: tables/table_evaluation_commercialmic.tex
\begin{table}[]
\footnotesize
\centering  
\caption{Evaluation on 7 microphone models. All of these microphones are vulnerable to~\alias and 6 of the 7 tested microphones can be injected clear voice commands.} 
\label{tab: eval_mic}
\setlength\tabcolsep{4pt} 
\renewcommand{\arraystretch}{1.2} 
\begin{tabular}{|c|c|c|cc|c|c|}
\hline
\multirow{2}{*}{Microphone   Model} & \multirow{2}{*}{Port} & \multirow{2}{*}{\begin{tabular}[c]{@{}c@{}}Auxiliary\\ Device\end{tabular}} & \multicolumn{2}{c|}{Parameters}  & \multirow{2}{*}{\begin{tabular}[c]{@{}c@{}}Inject\\ Audio\end{tabular}} & \multirow{2}{*}{L-dis} \\ \cline{4-5}
                                    &                            &                                                                             & \multicolumn{1}{c|}{fre.} & amp. &                                            &       \\ \hline
HUAWEI AM115                        & 3.5mm                      & Phone                                                                       & \multicolumn{1}{c|}{320}  & 220  & \Checkmark                                                                      & 0                   \\ \hline
HP DHP-1100l                        & 3.5mm                      & Phone                                                                       & \multicolumn{1}{c|}{30}   & 300  & \Checkmark                                                                       & 1                \\ \hline
Lenovo Lecoo MC01                   & 3.5mm                      & Phone                                                                       & \multicolumn{1}{c|}{315}  & 290  & \Checkmark                                                                     & 0               \\ \hline
UGREEN CM564                        & USB                        & Phone                                                                       & \multicolumn{1}{c|}{31}   & 280  & \Checkmark                                                                    & 13              \\ \hline
SM88                                & XLR                        & UFL-60                                                                      & \multicolumn{1}{c|}{320}  & 300  & \Checkmark                                                                     & 0              \\ \hline
TAKSTAR MS-118                      & XLR                        & UFL-60                                                                      & \multicolumn{1}{c|}{320}  & 260  & \Checkmark                                                                      & 0              \\ \hline
DS-KAU30HG-M              & XLR                        & UFL-60                                                                      & \multicolumn{1}{c|}{320}  & 250  & \Checkmark                                                                      & 0              \\ \hline
\end{tabular}
 \vspace{-0.7cm}
\end{table}

%% file: tables/table_evaluation_attack_distance.tex

\begin{table}[]
\footnotesize 
\centering  
\caption{\blue{Evaluation of attack distance}} 
\label{tab: attack distance}
\setlength\tabcolsep{7pt} 
\renewcommand{\arraystretch}{1.2} 
\begin{tabular}{|c|c|c|cccc|}
\hline
\multirow{2}{*}{\textbf{Test   Devices}} & \multirow{2}{*}{\textbf{Metrics}} & \multirow{2}{*}{\textbf{BL}} & \multicolumn{4}{c|}{\textbf{Attack Distance}}                                                                            \\ \cline{4-7} 
                                         &                                   &                              & \multicolumn{1}{c|}{\textbf{0.5m}} & \multicolumn{1}{c|}{\textbf{5m}} & \multicolumn{1}{c|}{\textbf{10m}} & \textbf{15m} \\ \hline
\multirow{2}{*}{\textbf{Microphone}}     & L-dis                             & 0                            & \multicolumn{1}{c|}{0}             & \multicolumn{1}{c|}{0}           & \multicolumn{1}{c|}{0}            & 0            \\ \cline{2-7} 
                                         & W2V                               & 0.79                         & \multicolumn{1}{c|}{0.70}          & \multicolumn{1}{c|}{0.73}        & \multicolumn{1}{c|}{0.69}         & 0.74         \\ \hline
\textbf{Camera}                          & Facenet                           & 99\%                         & \multicolumn{1}{c|}{97\%}          & \multicolumn{1}{c|}{97\%}        & \multicolumn{1}{c|}{98\%}         & 97\%         \\ \hline
\end{tabular}
 \vspace{-0.5cm}
\end{table}

%% file: sections/RelatedWork.tex

\section{\ \ \ Related Work}
\label{sec: relatedwork}
In this section, we introduce two types of injection attacks: wireless and wired injection attacks, that pose significant threats to sensor measurements. 

\subsection{Wireless Injection Attacks}
Extensive sensor manipulation studies over the past decades have focused on using wireless signals, including radiated EMI~\cite{tu2019trick, maruyama2019tap,wang2022ghosttouch, shan2022invisible,kohler2022signal, jiang2023glitchhiker,shoukry2013non, esteves2018remote, kasmi2015iemi,barua2020hall, dayanikli2020electromagnetic, dayanikli2022physical}, laser~\cite{sugawara2020light, jin2023pla}, sound~\cite{bolton2018bluenote, ji2021poltergeist}, and ultrasound~\cite{zhang2017dolphinattack}. For example, recent studies~\cite{maruyama2019tap,wang2022ghosttouch, shan2022invisible} show that radiated EMI can interfere with capacitance measurements, inducing false touches on touchscreens. Similarly, various sensors are susceptible to radiated EMI, including CCD sensors~\cite{kohler2022signal}, speed sensors~\cite{shoukry2013non}, temperature sensors~\cite{tu2019trick}, microphones~\cite{esteves2018remote, kasmi2015iemi}, hall sensors~\cite{barua2020hall}, IMU sensors~\cite{jang2023paralyzing}, etc. 
Additionally, Dayanikli et al.~\cite{dayanikli2022physical} utilized an IEMI signal to control the PWM-controlled actuators and previous works~\cite{kohler2022brokenwire, xu2005feasibility, jiang2023glitchhiker} have proposed that IEMI can disrupt wireless communication signals. 
The basic principle underlying these attacks is Faraday's law of induction~\cite{pearce1963faraday}, which states that a varying EM field can induce false currents in transmission cables. However, wireless EM signals are sharply attenuated as the distance from the source increases. Consequently, long-distance attacks often require enhanced hardware configurations, such as high-power microwave counter-drone weapons~\cite{droneweapon}.
Moreover, wireless EM attacks are easily mitigated by metal shielding enclosures, which are commonly used in IoT devices or sensors. In contrast,~\alias manipulates sensors via power cables, making it effective even for devices with heavy metal enclosures and at a long attack distance, such as 15\,m. 
Additionally, attackers can utilize laser, sound and ultrasound to spoof sensors, including lidars~\cite{jin2023pla}, cameras~\cite{sugawara2020light}, IMU~\cite{ji2021poltergeist}, microphones~\cite{zhang2017dolphinattack}, etc. While these attacks can achieve relatively long distances compared to radiated EMI-based attacks, their effectiveness is strictly limited by line-of-sight requirements, such as environmental visibility and obstructions between the signal source and the target device. On the contrary,~\alias is not constrained by these factors, as it exploits interconnected power cable to transmit attack signals, bypassing the need for direct visibility or physical proximity.

\subsection{Wired Injection Attacks}
Wired injection attacks, which exploit physical wires such as power cables to transmit attack signals and manipulate sensors, are an emerging threat to sensor measurements. 
For instance, Yang et al.~\cite{yang2023remote} showed that by injecting false currents into power lines, the switching mode power supply can emit sound, thereby spoofing nearby voice assistants. 
Similarly, Wang et al.~\cite{wang2022ghosttalk} presented that attackers could directly inject audio signals into a victim's phone via a modified charging cable.
Additionally, attackers~\cite{qiu2019voltjockey, selmane2008practical} have exploited power-supply manipulation to perform fault injection on digital circuits. For instance, Jiang et al.~\cite{jiang2022wight, jiang2023marionette} present wired GhostTouch attacks against capacitive touchscreens, where an attacker can induce fake touches on the capacitive touchscreen by injecting CM signals into the charging cable. However, these studies primarily focused on special sensors and lacked a comprehensive analysis of wired injection attacks, resulting in limited attack performance. 
A closely related attack to~\alias is Volttack~\cite{wang2023volttack}, which exploits power noise to modify the behavior of electronic components and interfere with sensors measurements. However, these wired attacks generally require compromising the power source and injecting differential voltages into the power cable, necessitating a more capable attacker compared to~\alias. Furthermore, differential attack signals are easily filtered by voltage stabilizers and attenuated by electronic components along the power cable. In comparison,~\alias relies on common-mode voltage by accessing the GND instead of differential voltages, which does not have such concerns.


This paper introduces a new attack threat to sensors, demonstrating how an attacker can manipulate sensor readings by injecting crafted signals into the interconnected GND at a distance, even enabling remote control of indoor sensors. We systematically studied the underlying principles of~\alias, established a general energy conversion model adaptable to a wide range of sensors, and identified the root causes of energy conversion through extensive modeling, simulations, and physical experiments. Furthermore, our work highlights an emerging threat from physical injection attacks, where an attack exploits the internal circuits of the victim device such as the GND wire as a potential antenna. This method exhibits the advantages of both radiated and conducted EMI, e.g., the interconnected GND not only extends the attack distance of radiated EMI but also eliminates the need for line-of-sight or close physical proximity. We believe the insights presented in this study shall guide for enhancing the sensor security and power wiring during the design phases.

%% file: sections/Discussion.tex

\section{Discussion}
\label{sec: disscussion}

\subsection{Potential Countermeasures}
~\alias exploits the GND wire to transmit voltages to the target sensor, leveraging internal vulnerabilities in sensor circuits to induce false measurements. However, since it relies on CM voltage rather than current, conventional power grid defenses such as circuit breakers and voltage stabilizers~\cite{scherz2006practical} are ineffective against~\alias, as demonstrated in~\mysec{sec: homegrid}. In this section, we propose potential countermeasures focusing on both attack detection and prevention strategies.



\subsubsection{Detection Methods} 

\begin{figure}[!t]  
	\centering
	\includegraphics[width=0.8\linewidth]{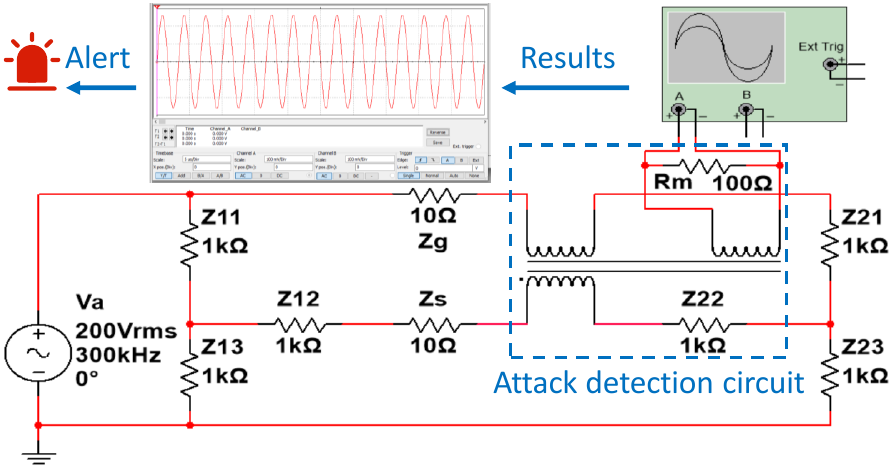}
	\caption{\blue{Illustration of the detection method. The simulated detection circuit consists of a 3P-CMC and a sensing resistance, which can detect the magnetic flux variation caused by~\alias.}}   
 \vspace{-1mm}
	\label{fig: CMdetection}
\end{figure}

To detect~\alias attacks and provide a protective response, we design a detection circuit based on a three-phase Common Mode choke (3P-CMC)~\cite{dominguez2018characterization}.
The 3P-CMC consists of three coils wound around a magnetic core, providing a low impedance path for differential signals and a high-impedance path for CM attack signals. Specifically, two windings are connected to the signal and GND cables of the device, while the third winding senses the magnetic flux variations caused by CM current in the GND and signal cables. To validate the effectiveness of this method, we conducted a simulation experiment as shown in~\fig{fig: CMdetection}, where $V_a$ represents the attack signal, and $R_m$ is the measuring resistance used to sense the induced voltage. The oscilloscope results show that the detection circuit successfully identifies the attack signal.


\subsubsection{Prevention Methods} To ensure sensor integrity under attack, we propose prevention methods. 

\textbf{Signal Attenuation.} The fundamental principle of~\alias involves injecting a CM signal into the GND to disrupt the victim device's output. To counter this threat, a straightforward and cost-efficient approach is to apply a signal filtering mechanism, such as adding a CM choke to the analog sensing circuit to eliminate CM current.


\textbf{Predictability Reduction.} The second defensive strategy is reducing adversaries' predictability by introducing random processing. Attackers often seek to mimic the victim device's output by forecasting its behavior. To counter this, designers are encouraged to implement random processing techniques that disrupt the predictability factor. For instance, to prevent an attacker from tuning the carrier frequency to induce a DC offset, as introduced in \mysec{sec: attack_signal_design}, designers can configure an ADC to sample at a random intervals. This can be achieved by adding a random delay, $t_{delay}$, to the original fixed sampling time $t_{k}$. The modified sampling time can then be expressed as $\hat{t}_{k} = t_{k} + t_{delay}$. This method is feasible and can be applied to various sensors. A similar randomized sampling-based prevention method has been validated in a prior study~\cite{trippel2017walnut}.

\textbf{Structure Optimization.} To mitigate the threat of~\alias, an effective way is to cut off the coupling path, specifically the long analog signal wire and the GND wire of the sensor's output. This can be achieved by adopting soldering techniques or shorting the analog signal paths. Since the asymmetry of circuit impedance is the root cause of CM-DM conversion, a well-designed Printed Circuit Board (PCB) with a symmetrical layout and balanced component placement can effectively prevent CM noises from being converted into disruptive DM signals.

%% file: sections/Conclusion.tex
\section{\ Conclusion}
\label{sec: conclusion}
we present~\alias, a new attack vector that manipulates sensor readings remotely by injecting signals into the GND wire, enabling cross-room and cross-socket attacks. Through an in-depth analysis of energy conversion principles and root causes, we validate~\alias on 17 off-the-shelf sensors and demonstrate its effectiveness in real-world attacks on a surveillance system and a broadcast system. Additionally, we establish a home power system to evaluate the impact of electrical factors on~\alias and propose countermeasures to mitigate its threat.

%% file: sections/Acknowlegment.tex
\section*{Acknowledgment}
We thank the anonymous shepherd and reviewers for their valuable comments. This work is supported by the National Natural Science Foundation of China (NSFC) Grant 62222114, 61925109, and 62071428.

%% file: sections/Appendix.tex

\newpage
\appendix
\section{Appendix}\label{sec: appendix}

\subsection{\textsc{Modeling Simulation Parameters}}\label{sec: appendix_modeling_simulation}
The simulation parameters of the coupling model and conversion model are defined as follows.

\noindent(1) Simulation Parameters in Coupling Stage.
\begin{equation}{
    \begin{aligned}
    V_s &= 300 Vpp \\[1pt]
    Z_{ga1} &= 1000000+\frac{1}{j\times 1\times 10^{-5} \omega } \Omega \\[1pt]
    Z_{sa1} &= 1000000+\frac{1}{j\times 0.99\times 10^{-5} \omega }\Omega \\[1pt]
    Z_{ga2} &= 1000000+\frac{1}{j\times 1.01\times 10^{-5} \omega }\Omega \\[1pt]
    Z_{sa2} &= 1000000+\frac{1}{j\times 0.98\times 10^{-5} \omega }\Omega \\[1pt]
    Z_{gs1} &= 1000000+\frac{1}{j\times 1.21\times 10^{-5} \omega }\Omega \\[1pt]
    Z_{gs1} &= 1000000+\frac{1}{j\times 1.19\times 10^{-5} \omega }\Omega \\[1pt]
    Z_{s} &= 0.00099+j\times 5\times 10^{-6} \omega+\frac{1}{j\times 1.1\times 10^{-9} \omega }\Omega \\[1pt]
    Z_{g} &= 0.0001001+j\times 4.43\times 10^{-6} \omega+\frac{1}{j\times 0.99\times 10^{-9} \omega }\Omega
    \end{aligned}\label{eq: simulation_coupling}
    \nonumber}
\end{equation}

\noindent(2) Simulation Parameters in Converting Stage.
\begin{equation}{
    \begin{aligned}
    Z_{1I} &= 1000000+\frac{1}{j\times 1\times 10^{-7} \omega }\Omega \\[1pt]
    Z_{2I} &= 1000000+\frac{1}{j\times 1.1\times 10^{-7} \omega }\Omega \\[1pt]
    Z_{1O} &= 1000000+\frac{1}{j\times 0.99\times 10^{-7} \omega }\Omega \\[1pt]
    Z_{2O} &= 1000000+\frac{1}{j\times 1.01\times 10^{-7} \omega }\Omega \\[1pt]
    Z_{3I} &= 1000000+\frac{1}{j\times 1.21\times 10^{-6} \omega }\Omega \\[1pt]
    Z_{3O} &= 1000000+\frac{1}{j\times 1.19\times 10^{-6} \omega }\Omega \\[1pt]
    Z_{R} &= 19.99+j\times 0.05 \omega+\frac{1}{j\times 1.1\times 10^{-3} \omega }\Omega \\[1pt]
    Z_{L} &= 20.01+j\times 0.049 \omega+\frac{1}{j\times 1.2\times 10^{-3} \omega }\Omega
    \end{aligned}
    \nonumber
    }
\end{equation}

\subsection{\textsc{Supplementary Materials of Evaluation}}\label{sec: supplement_eval}


\begin{figure}[h]
	\centering
	\includegraphics[width=0.9\linewidth]{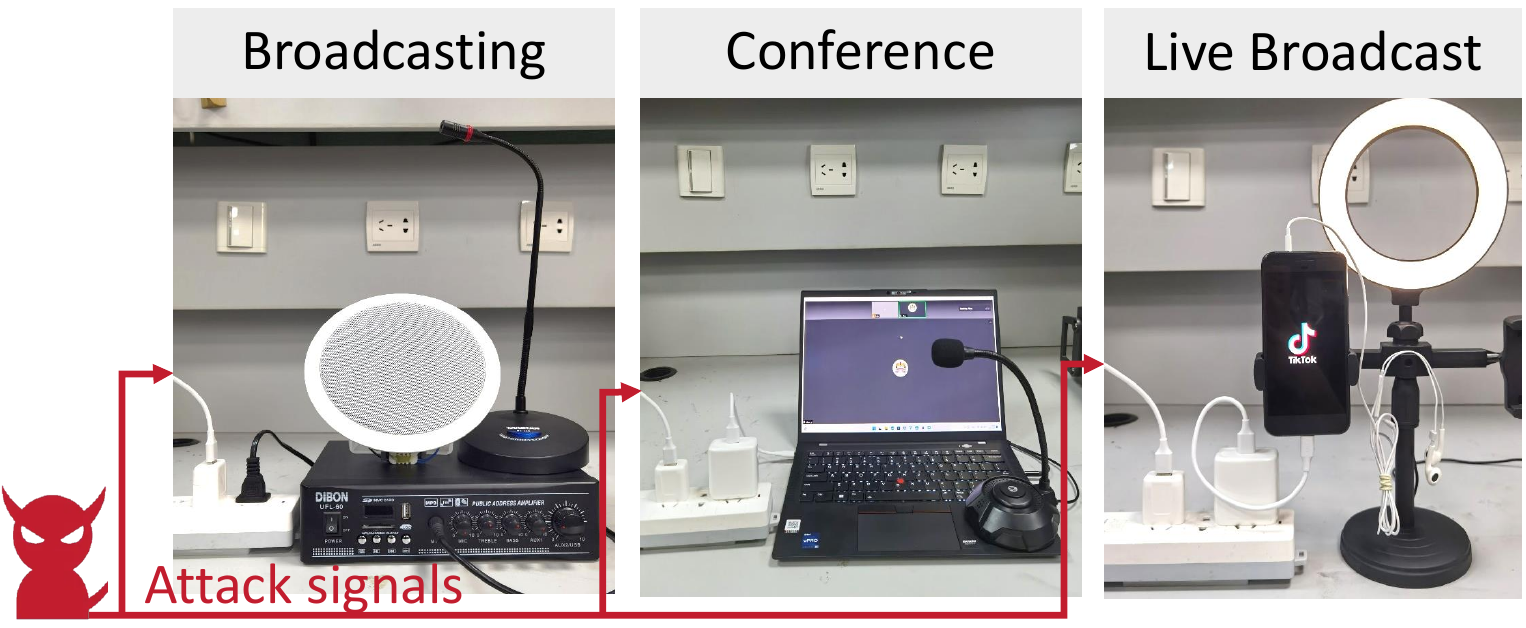}
	\caption{Illustrations of microphone attack scenarios. The attacker can successfully inject audio into the public broadcasting system, remote conference system, and live broadcasting system.  }
	\label{fig: setup_mic}
\end{figure}

\begin{figure}[h]
    \centering
    \includegraphics[width=0.9\linewidth]{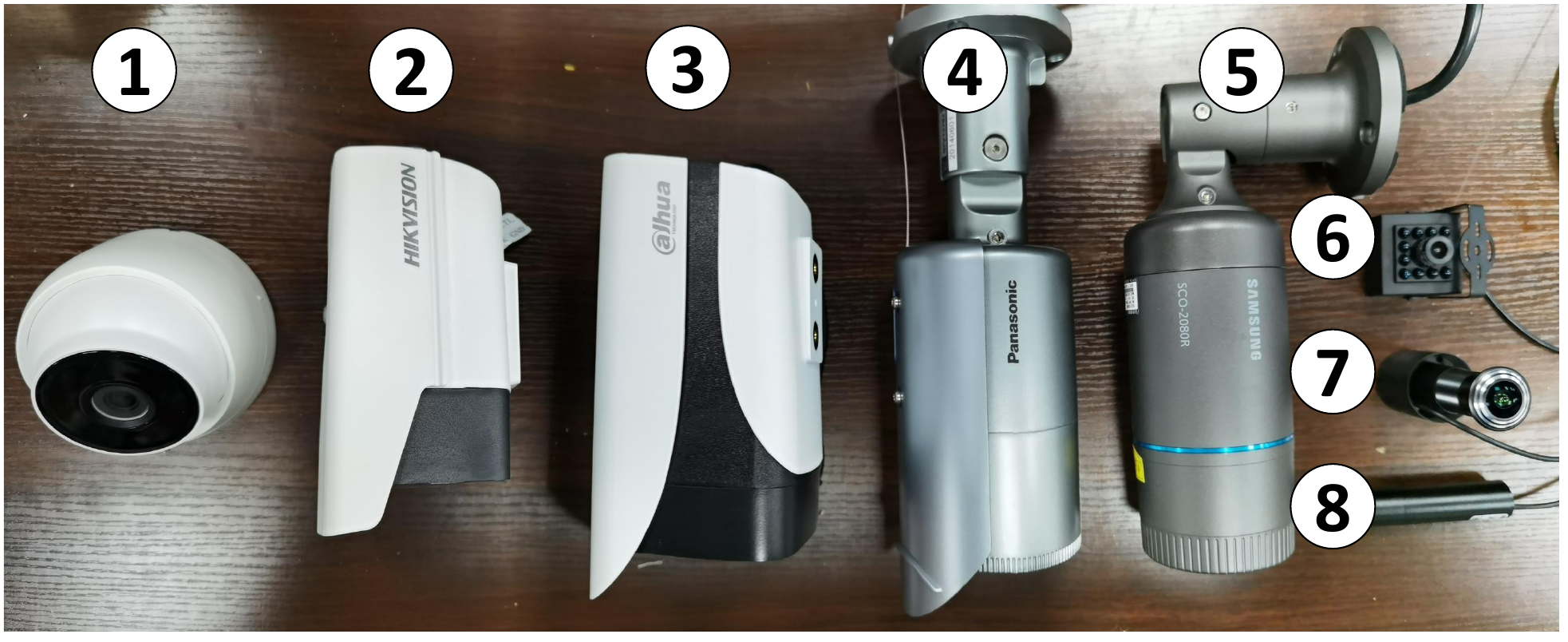}
    \caption{The 8 commercial analog cameras tested in the evaluation: \ding{172} HIKVISION DS-2CE56C3T-IT3~\cite{hikvisiono2024ds2ce56cot}, \ding{173} HIKVISION DS-2CE16G0T-IT3~\cite{hikvisiono2024ds2ce16d0t}, \ding{174} DH-HAC-HFW1200M-I2~\cite{dahua2024dhhac}, \ding{175} Panasonic WV-CW314LCH~\cite{panasonic2024wvcw314lCH}, \ding{176} SAMSUNG SCO-2080RP~\cite{samsung2024sco2080r}, \ding{177} SONY CCD673-1200, \ding{178} SONY CCD-1200, and \ding{179} SONY IMX323 ~\cite{sony2024imx323}.}
    \label{fig: eval_camera_list}
\end{figure}

\begin{figure}[h]
    \centering
    \includegraphics[width=0.9\linewidth]{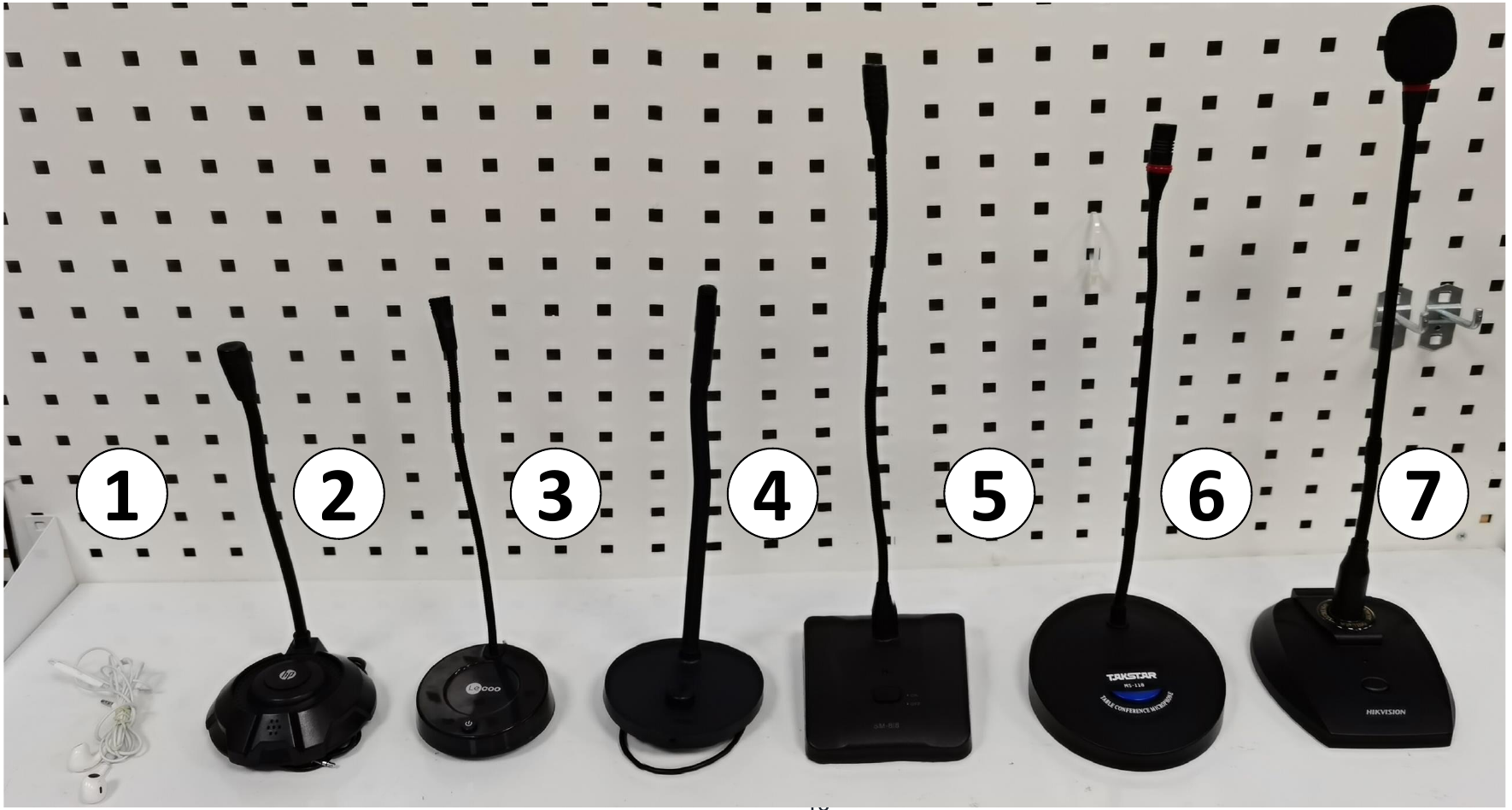}
    \caption{The 7 commercial microphones tested for evaluation: \ding{172} wired earphone HUAWEI AM115~\cite{productz2024huaweimic}, \ding{173} desktop microphone HP DHP-1100l~\cite{jd2024hpmic}, \ding{174} desktop wired microphone Lenovo Lecoo MC01 ~\cite{banggood2024lenovmic}, \ding{175} USB microphone UGREEN CM564~\cite{jg2024ugreen}, \ding{176} broadcast microphone SM88, \ding{177} conference microphone TAKSTAR MS-118~\cite{takstar2024ms118}, \ding{178} conference microphone HIKVISION DS-KAU30HG-M~\cite{jd2024ds-kau30hg-m}.}
	\label{fig: eval_mic_list}
\end{figure}

\begin{figure}[h]  
	\centering
	\subfigure[Target voice signal.]{
		\includegraphics[width=0.45\linewidth]{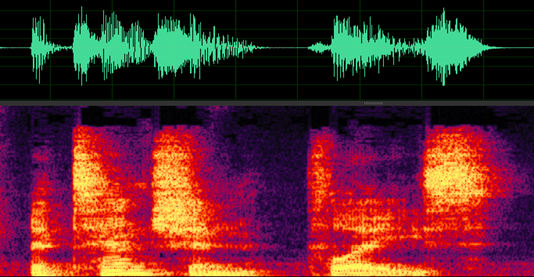}
		\label{fig: original_audio}
	}
	\subfigure[Injected voice signal.]{
		\includegraphics[width=0.45\linewidth]{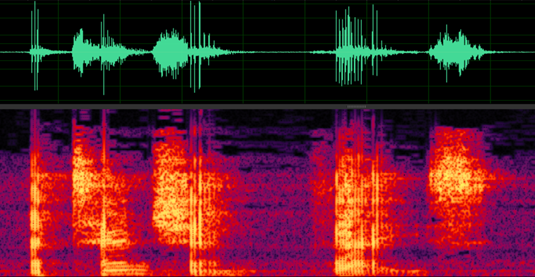}
		\label{fig: injected_audio}
	}
 \caption{Recordings and spectrograms of (a) the target voice audio and the (b) injected voice audio (``Attention, please").}
 \label{fig: results_audio}	
\end{figure}

\begin{figure}[!t]  
    \centering
    \subfigure[Back panel.]{\includegraphics[height=0.45\linewidth]{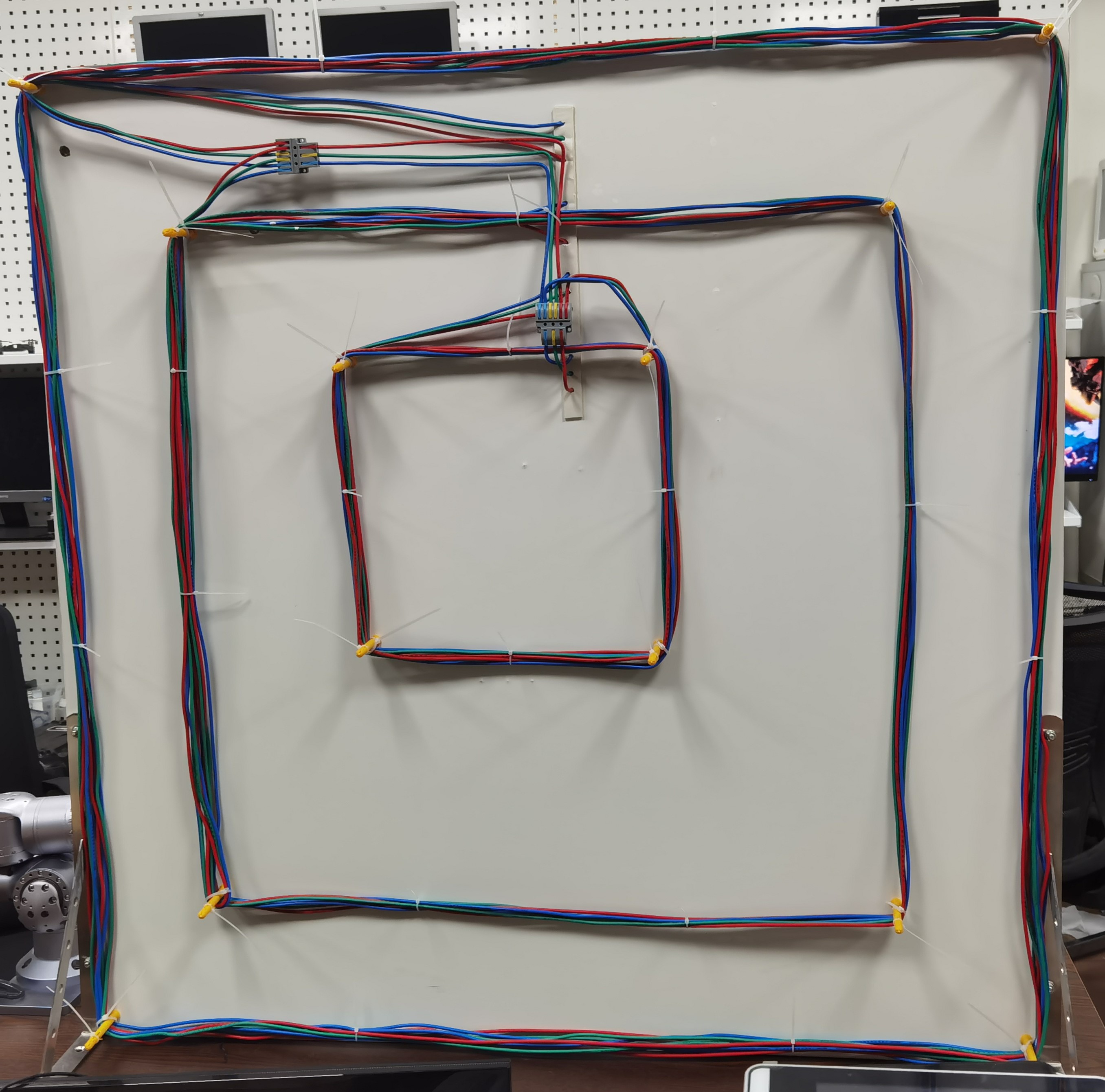}\label{fig: back panel}
	}\hspace{-0ex}
    \subfigure[Part of grid.]{
	\includegraphics[height=0.45\linewidth]{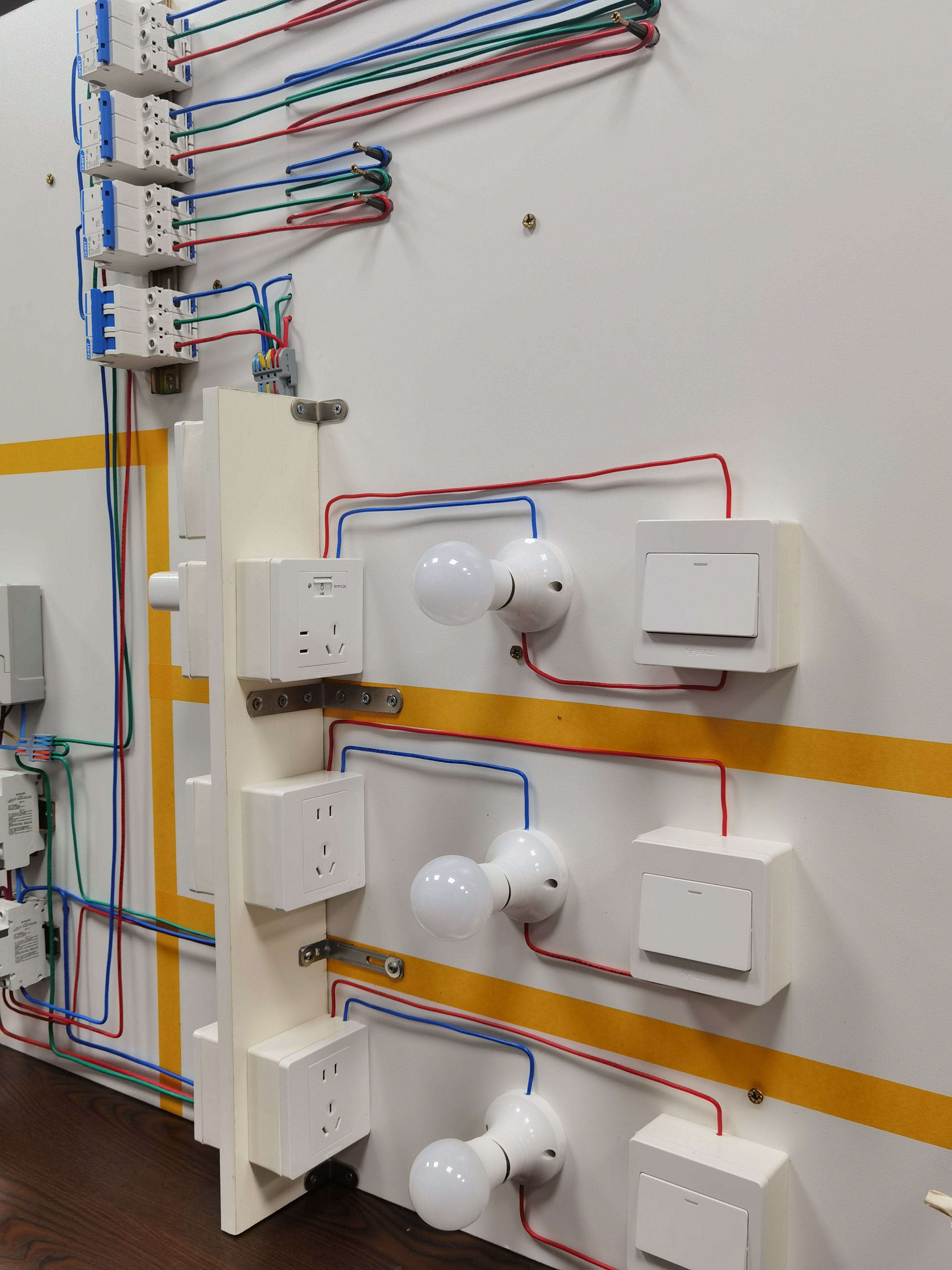}\label{fig: part of grid}
	}
    \caption{\blue{Illustration of the back panel of the household power distribution system (left) and the side view of the system (right).}}
\vspace{-1cm}
\label{fig: backpanel_and_grid}
\end{figure}

\begin{figure}[!t]
    \centering
    \includegraphics[width=0.9\linewidth]{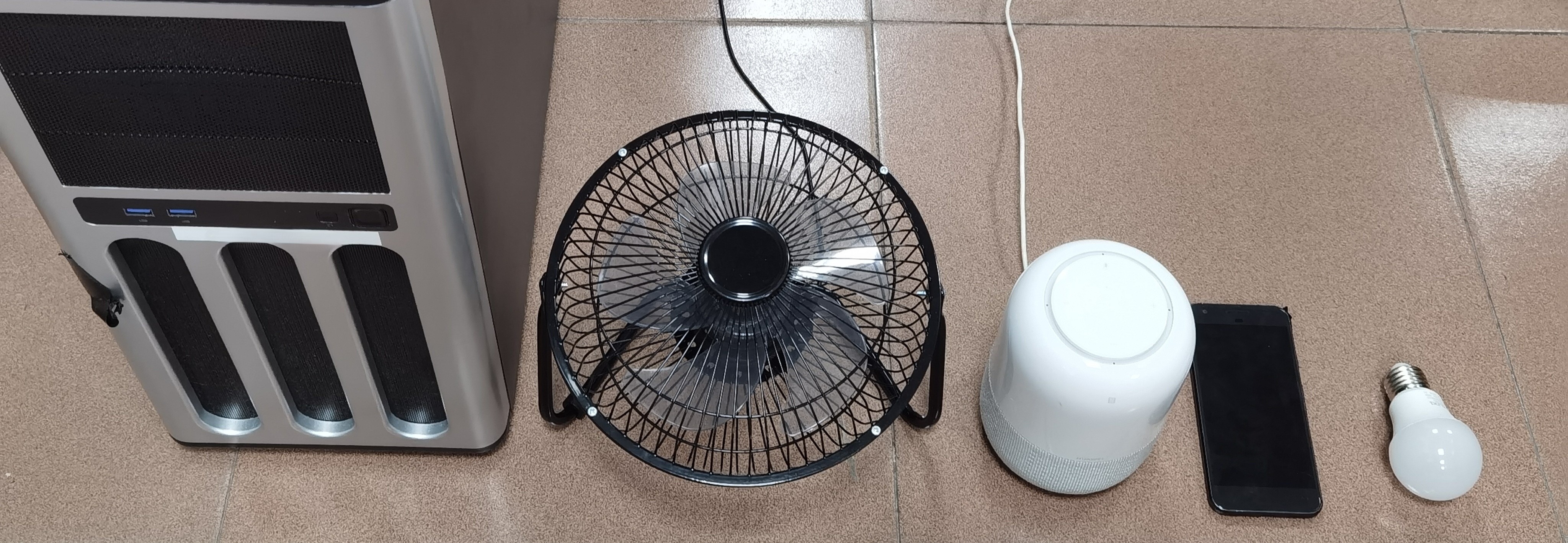}
    \caption{We evaluate the impact of electrical noises on~\alias by plugging 5 typical IoT devices into the household power system.}
    \label{fig: microgrid_3}
    \vspace{-1cm}
\end{figure}

\begin{figure}[h]  
	\centering
	\subfigure[Setup.]{
\includegraphics[width=0.48\linewidth]{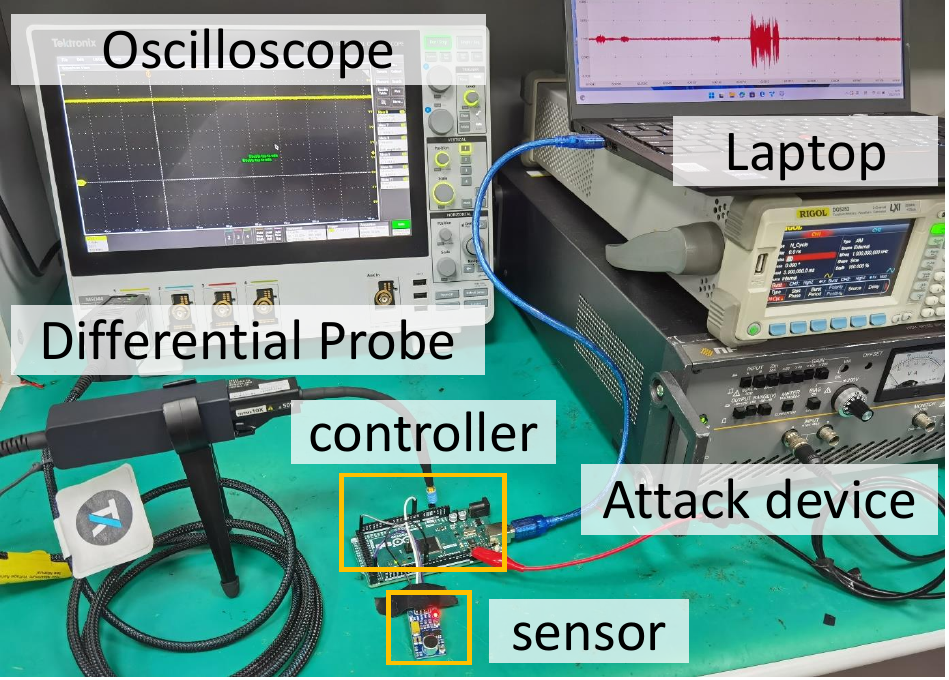}
		\label{fig: tivp02_setup}
	}
    \hspace{-0.5em}
	\subfigure[Results.]{\includegraphics[width=0.47\linewidth]{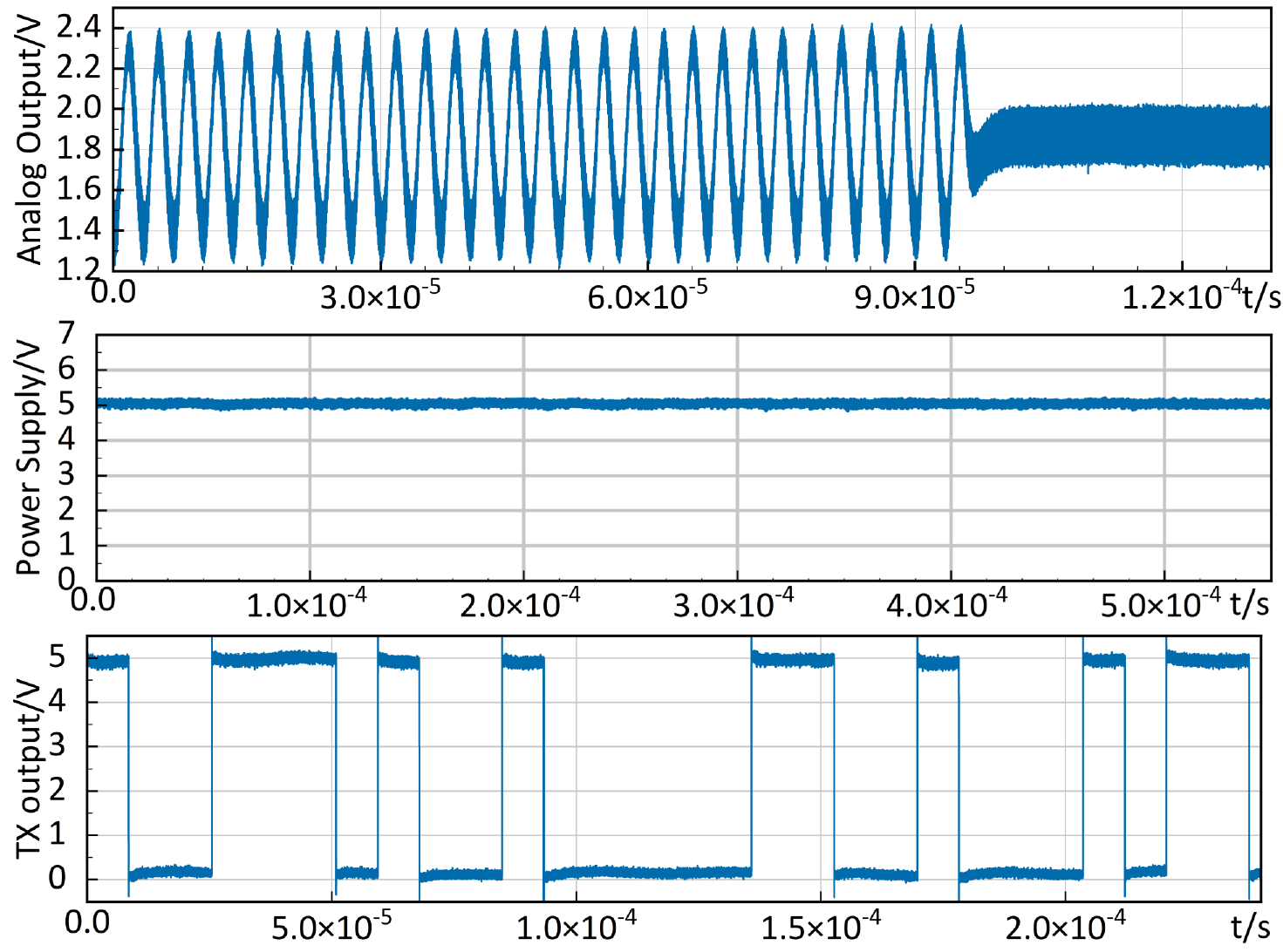}
		\label{fig: dm_probe_test}
	}
	\caption{Experiment of effect on driver board. (a) Experimental setup of measuring the output of the driver board under attack by using the differential high-voltage probe. (b) Results of the differential probe. Top: analog output of the tested module. Middle: power supply voltage of the driver board. Bottom: digital communication of the laptop.}
    \label{fig: attacksignal_design}
\end{figure}